\documentclass[twocolumn]{aastex63}

\received{}
\revised{}
\accepted{}
\submitjournal{ApJ}

\newcommand{\C}[1]{{\color{cyan} \bf #1}}
\newcommand{\R}[1]{{\color{red} \bf #1}}
\newcommand{\N}[1]{{\color{black} \bf #1}}

\shorttitle{SN 2015bq}
\shortauthors{Li et al.}

\begin{document}
	
	\title{\textbf{SN 2015bq: A Luminous Type Ia Supernova with Early Flux Excess}}

	\author{Liping Li}
	\affiliation{School of Physics and Astronomy, Yunnan University, Kunming 650091, China}
	\affiliation{Yunnan Observatories (YNAO), Chinese Academy of Sciences, Kunming 650216, China;  jujia@ynao.ac.cn.}

	\author{Jujia Zhang*}
	\affiliation{Yunnan Observatories (YNAO), Chinese Academy of Sciences, Kunming 650216, China;  jujia@ynao.ac.cn.}
	\affiliation{Key Laboratory for the Structure and Evolution of Celestial Objects,CAS, Kunming, 650216, China}
	
	\author{Benzhong Dai}
	\affiliation{School of Physics and Astronomy, Yunnan University, Kunming 650091, China}
	
	\author{Wenxiong Li}
	\affiliation{The School of Physics and Astronomy, Tel Aviv University, Tel Aviv 69978, Israel}	

	\author{Xiaofeng Wang}
	\affiliation{Physics Department and Tsinghua Center for Astrophysics (THCA), Tsinghua University, Beijing, 100084, China}	
	
	\author{Qian Zhai}
	\affiliation{Yunnan Observatories (YNAO), Chinese Academy of Sciences, Kunming 650216, China;  jujia@ynao.ac.cn.}
	\affiliation{Key Laboratory for the Structure and Evolution of Celestial Objects,CAS, Kunming, 650216, China}
	
	\author{Jinming Bai}
	\affiliation{Yunnan Observatories (YNAO), Chinese Academy of Sciences, Kunming 650216, China;  jujia@ynao.ac.cn.}
	\affiliation{Key Laboratory for the Structure and Evolution of Celestial Objects,CAS, Kunming, 650216, China}
	
	
	
	\begin{abstract}
		We present optical and ultraviolet (UV) observations of a luminous type Ia supernova (SN Ia) SN 2015bq characterized by the early flux excess. This SN reaches a B-band absolute magnitude at $M_B = -19.68 \pm 0.41$ mag and a peak bolometric luminosity at $L = (1.75 \pm 0.37) \times 10^{43}$ erg s$^{-1}$, with a relatively small post-maximum decline rate [$\Delta m_{15}(B) = 0.82 \pm 0.05$ mag]. 
		The flux excess observed in the light curves of SN 2015bq a few days after the explosion, especially seen in the UV bands, might be due to the radioactive decay of $^{56}$Ni mixed into the surface. The radiation from the decay of the surface $^{56}$Ni heats the outer layer of this SN. It produces blue $U-B$ color followed by monotonically reddening in the early phase, dominated iron-group lines, and weak intermediate-mass elements absorption features in the early spectra. The scenario of  enhanced $^{56}$Ni in the surface is consistent with a large amount of $^{56}$Ni ($M_{ \rm ^{56}{\rm Ni}}$ = 0.97 $\pm 0.20$ $M_{\sun}$) synthesized during the explosion. The properties of SN 2015bq are found to locate between SN 1991T and SN 1999aa, suggesting the latter two subclasses of SNe Ia may have a common origin.

	\end{abstract}
	
	
	
	\section{Introduction} 
	Type Ia supernovae (SNe Ia) are  excellent distance indicators on cosmic scales whose peak luminosities can be well  calibrated by the relation with the width of the light curve \citep[i.e., width-luminosity relation, WLR;][]{1993ApJ...413L.105P}. 
	Moreover, the studies of SNe Ia provide the first evidence for  accelerating expansion of the universe \citep{1998AJ....116.1009R,1999ApJ...517..565P}, indicating that the current universe is dominated by dark energy. 
	Current research shows that SNe Ia are produced by thermonuclear explosion of a carbon-oxygen white dwarf (WD) in the binary system when approaching the Chandrasekhar-mass  \citep[$\sim$ 1.38$M_{\sun}$;][]{2000ARA&A..38..191H,2014ARA&A..52..107M}. At the same time, the sub-Chandrasekhar model has also been examined to account for most SNe Ia \citep[e.g.,][]{2000tias.conf...33L,2010ApJ...714L..52S,2020MNRAS.491.2902F}.
	There are two main progenitor scenarios. One is the single-degenerate (SD) scenario, in which the WD reaches the Chandrasekhar-mass limit by accreting matter from a nondegenerate companion star \citep{1973ApJ...186.1007W,1982ApJ...253..798N,1997Sci...276.1378N}. 
	Another is the double-degenerate (DD) scenario, where two WDs in the close binary system eventually merge and explode by losing energy and angular momentum through gravitational wave radiation \citep{1984ApJS...54..335I,1984ApJ...277..355W}.
	
	According to the spectroscopic characteristics, SNe Ia could be divided into several subclasses \citep{2005ApJ...623.1011B,2009PASP..121..238B,2009ApJ...699L.139W}. For example, SNe Ia characterized by relatively weak absorption of \ion{Si}{2} $\lambda$6355 and \ion{Si}{2} $\lambda$5972 are divided into shallow silicon (SS) subclass in the scheme of \cite{2006PASP..118..560B,2009PASP..121..238B}.
	
	However, SS is not a homogeneous category.  It includes the SNe Ia possibly originating from a possible super-Chandrasekhar-mass progenitor, i.e., SN 2003fg \citep{2006astro.ph..9804J}, SN 2006gz \citep{2007ApJ...669L..17H}, SN 2007if \citep{2010ApJ...713.1073S}, SN 2009dc \citep{2011MNRAS.412.2735T}, the luminous 1991T/1999aa-like events dominated by higher ionization lines (\ion{Fe}{3}) at the early phase,  the narrow-lined events (NL, e.g., SN 2012fr \citep{2014AJ....148....1Z}) located between 91T and the normal ones,  the spectroscopically normal events (e.g., SN 2006S \citep{2012AJ....143..126B}), and the faint 2002cx-like SNe \citep{2003PASP..115..453L,2006PASP..118..722C,2014ApJ...786..134M}. 
	
	The dense and timely observations are essential to understand the physical origin of the diversity seen in the SS SNe Ia. This paper presents such data for SN 2015bq, a luminous 99aa-like event that shows flux excess at the early phase.
	Based on the optical and ultraviolet observations of SN 2015bq, we analyze its light curves and the color curves in Section~\ref{sec:3}. The investigations of the spectra are presented in Section~\ref{sec:4}.
	Next we discuss the mass of $^{56}{\rm Ni}$, the early excess and the diversity among luminous SS Ia in Section \ref{sec:5}. Finally, a summary is given in Section~\ref{sec:6}.	
	
	\section{Observations and Data Reduction}\label{sec:2}
\begin{figure}
		\includegraphics[width=\columnwidth]{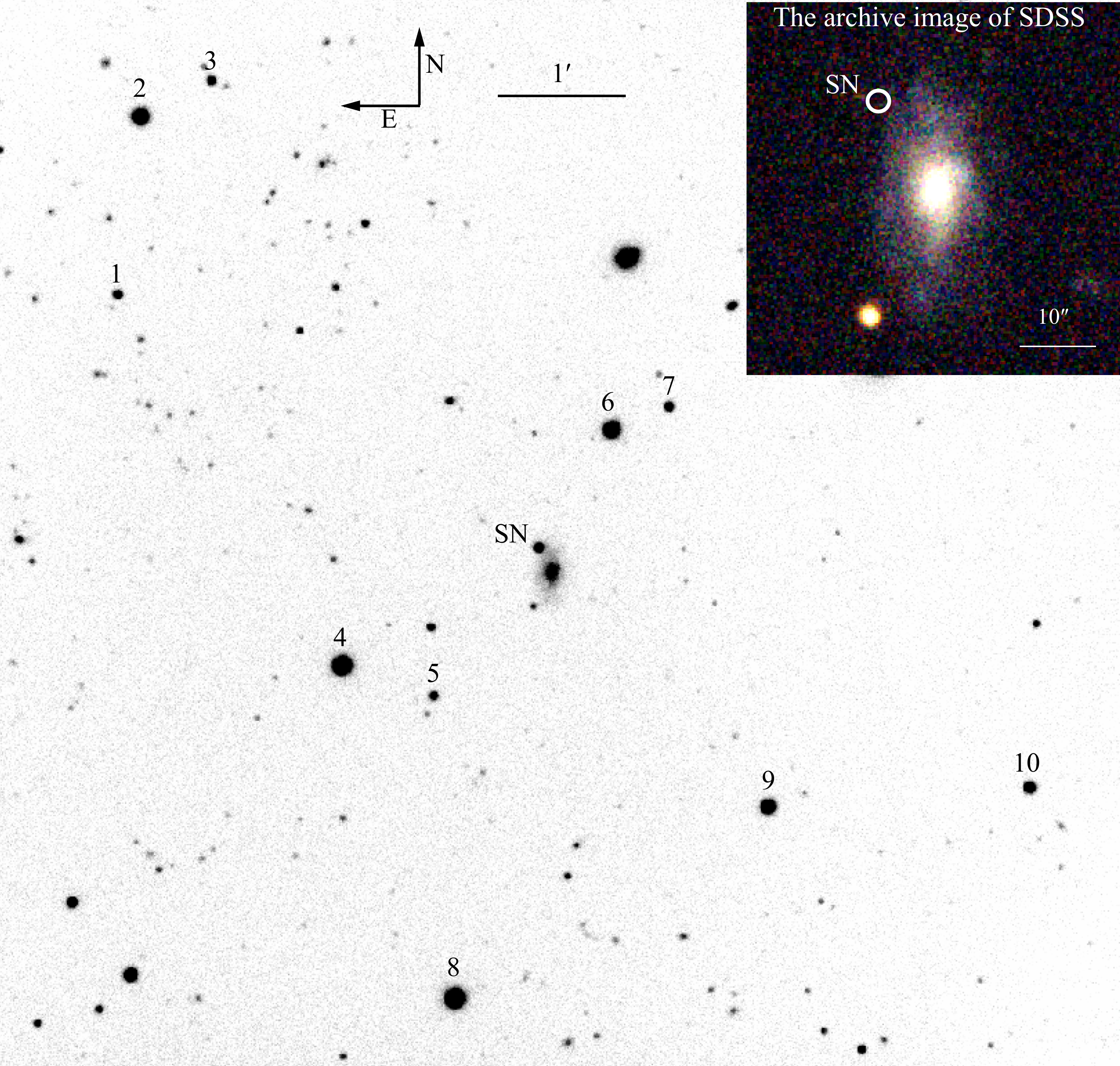}
		\caption{Finder chart of SN 2015bq in LEDA 41898, taken by the LJT. The supernova and ten local reference stars are marked. The archival data of the host galaxy combined from the gri band images of SDSS is plotted at the top-right. \label{fig:img}}
	\end{figure}

	SN 2015bq was discovered independently by four transient survey projects \citep[PSN J12350637+3114354 = PS15xn = CSS150219:123506+311436 = iPTF15ku;][]{2015ATel.7119....1F}. Its coordinates are RA = {12$^h$}{35$^m$}{06$\fs$37}, DEC = +{31$\degr$}{14$\arcmin$}{35$\farcs$4} (J2000), located at {71$\arcsec$} west and {15$\farcs$5} north of the center of the host galaxy LEDA 41898, which has a redshift of 0.028 \citep{1985AnTok..20..335T} (see Figure \ref{fig:img}). 
	
	Tsinghua University-NAOC Transient Survey \citep[TNTS;][]{2015RAA....15..215Z} initially reported the discovery of SN 2015bq via an unfiltered image at about 17.9 mag on February 16.72, 2015 UT (Universal Time is used throughout this paper). However, this transient was also independently discovered at about 2.3 days earlier by Palomar Transient Factory (PTF) at roughly a magnitude of PTF-g = 18.5 mag on February 14.43 2015 \citep{2015ATel.7119....1F}. 
	
	It was classified as a young 99aa-like event based on the spectra taken by Lijiang 2.4 meter telescope \citep[LJT;][]{2015RAA....15..918F} with Yunnan Faint Object Spectrograph and Camera \citep[YFOSC;][]{2019RAA....19..149W} on February 18.9 \citep{2015ATel.7109....1Z}, and Nordic Optical Telescope (NOT) with Andalucia Faint Object Spectrograph and Camera (ALFOSC) on February 15 \citep{2015ATel.7119....1F}, respectively. 
	
	Thanks to the supernova program (LiONS, Lijiang One hour per Night for Supernova observation) of LJT, we can monitor this transient frequently at LJT in both photometric and spectroscopic modes. The observing campaign at LJT spanned from $t \approx  -$12 d to $t \approx  +$106 d relative to the $B$-band maximum ($ t $ is the time since $B$-band maximum and is used throughout this paper). 
	Meanwhile, The Tsinghua-NAOC 0.8 m telescope  \citep[TNT;][]{2008ApJ...675..626W,2012RAA....12.1585H}  and Xing-Long 2.16 m telescope (XLT) at Xing-long Observation of National Astronomical Observatories (NAOC) also participated in the optical photometric and spectroscopic monitoring, respectively. Furthermore, the UV/optical data of SN 2015bq were also collected by the Ultraviolet/Optical Telescope \citep[UVOT;][]{2005SSRv..120...95R} onboard the \textit{Swift} satellite also started, i.e., starting at $t \approx $ -15 d. 
		
	\subsection{Photometry Observation}
	Nearly daily \textit{UBVRI}-band photometry of SN 2015bq were conducted at LJT for the first two months after the discovery. The ground-based optical photometry of SN 2015bq obtained by the LJT and TNT covered the period from t $ \approx -12$ d to t $ \approx +106$ d. 
	The CCD images were reduced using the IRAF \footnote{IRAF,the Image Reduction and Analysis Facility, is distributed by the National Optical Astronomy Observatory, which is operated by the Association of Universities for Research in Astronomy (AURA), Inc. under cooperative agreement with the National Science Foundation (NSF).} standard procedure, including bias subtraction, flat fielding, and removal of cosmic rays. 
	We performed background subtraction of the host galaxy light for all filters using template observations gathered in February 2016. Aperture photometry was applied to the image after template subtraction. 
	The instrumental magnitudes of SN 2015bq are further converted to the standard Johnson UBV and Kron-Cousins RI systems based on the ten local standard stars (as labeled in Figure \ref{fig:img} and listed in Table \ref{tab:stand}). 
	The final results of the photometry from the LJT and TNT are listed in Table \ref{tab:Opti_Pho}.
	
	The \textit{Swift}-UVOT observations have covered for about the first month in three UV filters (uvw2, uvm2, and uvw1) and three broadband optical filters (u, b, and v). 
	The UV-Optical photometry data presented in Table \ref{tab:UVOT} has been published on the Swift Supernovae \footnote{http://people.physics.tamu.edu/pbrown/SwiftSN/swift\_sn.html} website by Peter Brown. The UVOT data are reduced using the Swift's Optical/Ultraviolet Supernova Archive \citep[SOUSA;][]{2014Ap&SS.354...89B} reductions, including subtracting the underlying host galaxy using \textit{Swift}-UVOT observations from October and November 2016. 
	Note that the upper limits of the uvm2 band are listed in Table \ref{tab:UVOT} since there are no reliable detections of SN 2015bq in this band.
	
	\subsection{Spectroscopic Observation}
	Twenty-three low-resolution spectra collected by the LJT, spaning from t $= -$10 d to t = +83 d, are listed in Table 4 and presented in Figure 2.  Furthermore, one spectrum was also obtained with the BFOSC (Beĳing Faint Object Spectrograph and Camera) mounted on the Xinglong 2.16-m telescope.  
	
	All spectra were reduced using standard IRAF routines. The spectra were calibrated with the spectrophotometric  standard stars observed at a similar air mass on the same night. Furthermore, the spectra were corrected for the continuum atmospheric extinction at the Lijiang Observatory, and the telluric lines were also removed.
	
\begin{figure}
		\includegraphics[width=\columnwidth]{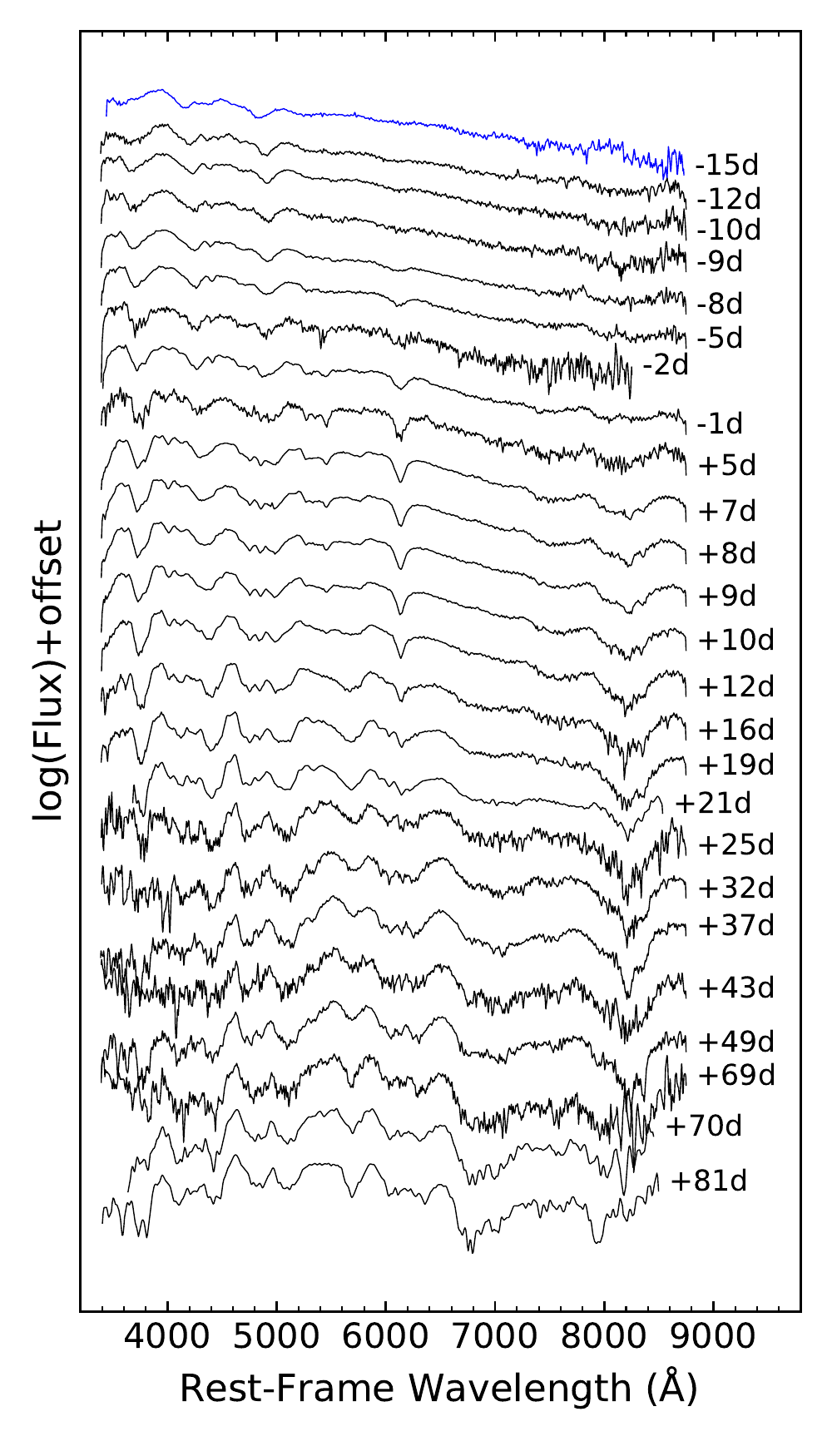}
		\caption{Optical spectral evolution of SN 2015bq. The spectra have been corrected for the redshift of the host galaxy ($ v_{hel} = 8448\, \rm km\, s^{-1}$) and telluric lines. Note the first spectra from NOT in blue. The numbers on the right-hand side mark the epochs of the spectra in days after the $B$-band maximum. (Supplemental data for this figure are available in the online journal).  \label{fig:Sp}}
	\end{figure}

	Besides, a spectrum obtained on Feb. 05, 2015, at NOT with ALFOSC \citep{2015ATel.7119....1F} presented at WISeREP \footnote{http://wiserep.weizmann.ac.il/spectra/list} \citep{2012PASP..124..668Y} is also plotted in Figure \ref{fig:Sp} for further analysis and discussion. In addition, the continuums of these spectra have been checked with the UBVRI-band photometry at similar phases.

	\section{Photometry}\label{sec:3}
	\subsection{Optical and Ultraviolet Light Curves}
\begin{figure}
		\includegraphics[width=\columnwidth]{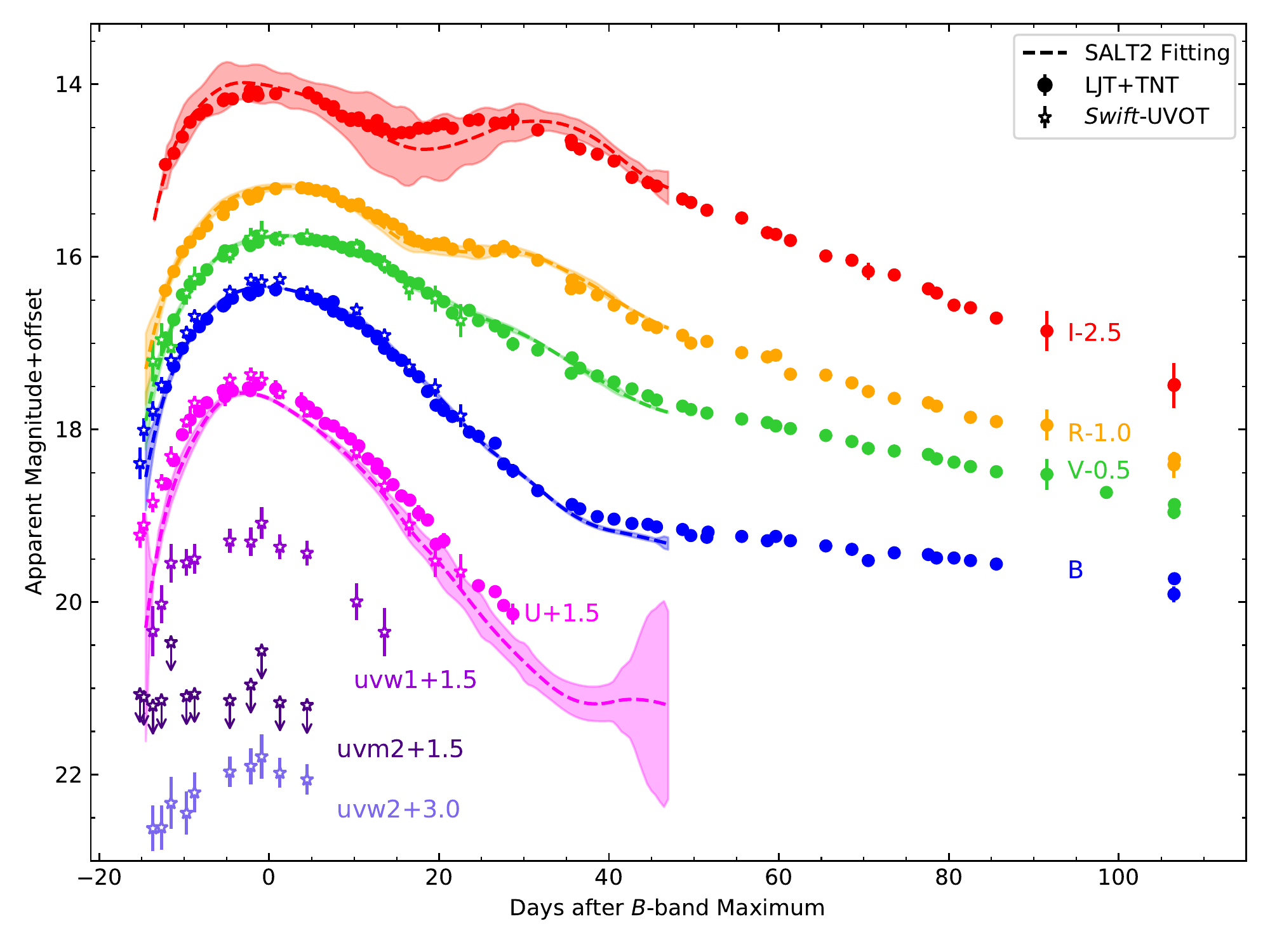}
		\caption{Ultraviolet and optical light curves of SN 2015bq. The upper limits of the non-detection of SN 2015bq in the uvm2 band are plotted with the downside arrow . The light curves are shifted vertically for better display. The dashed-lines are the SALT2 fitting \citep{2007A&A...466...11G}. \label{fig:LC}}
	\end{figure}

	The optical and UV light curves of SN 2015bq are displayed in Figure \ref{fig:LC}.
	
	Using a polynomial fit to the observed light curves, we derived the parameters of peak magnitudes, the dates at the maximum brightness, and light-curve decline rates, as listed in Table \ref{tab:lc_param}. 
	SN 2015bq reaches the $B$-band peak magnitude at $B_{\max}$ = 16.39 $\pm$ 0.01 mag on JD 2457084.62 $\pm$ 0.30, with the corresponding decline rate as $\Delta$$m_{15}(B)$ = 0.78 $\pm$ 0.03 mag. According to the width-luminosity relation, the small decline rate means that SN 2015bq has a high peak luminosity and synthesizes a significant amount of $^{56}{\rm Ni}$. 
	The prominent $I$-band and $R$-band secondary maximum of SN 2015bq might be due to the Fe emission from high excitation transitions following recombination of doubly ionized elements to singly ionized states, usually seen in the luminous SNe Ia \citep{2015MNRAS.449.3581J,2015ApJ...813...30S}. It is consistent with strong \ion{Fe}{3} lines in the early spectra and  amount of $^{56}$Ni estimated in Section 3.3.

	Figure 3 displays the SALT2 \citep[the spectral adaptive light curve template;][]{2007A&A...466...11G} fitting of the optical light curves of SN 2015bq. The fitting parameters and results are listed in Table \ref{tab:S2_param}, in which the maximum brightness matches the results from the polynomial ﬁt. 
	We calculate the $\Delta$$m_{15}(B)$, extinction, and distance modulus using those parameters \citep{2018PASP..130f4101V},  which yields $\Delta$$m_{15}(B)$ as 0.86 $\pm$ 0.06 mag, $E(B - V)$ as 0.15 $\pm$ 0.07 mag and $\mu_{0}$ as 35.55 $\pm$ 0.14 mag.
	We take the average value $\Delta$$m_{15}(B)$ as 0.82 $\pm$ 0.05 mag from the above methods.
		
	\begin{deluxetable}{ccccc}
		\tablecaption{Light-curve Parameters of SN 2015bq \label{tab:lc_param}}
		\tablehead{
			\colhead{Band} & \colhead{$\lambda_{eff}$} & \colhead{$t_{max}$\tablenotemark{a}} & \colhead{$m_{peak}$\tablenotemark{b}} & \colhead{$\Delta m_{15}$\tablenotemark{b}} \\
			\colhead{ } & \colhead{(\AA)} & \colhead{(JD-2457000.5)} & \colhead{(mag)} & \colhead{(mag)} 
		}
		\startdata
		uvw2 & 1928 & 83.23(40) & 18.80(26) & .. \\
		uvw1 & 2600 & 83.23(40) & 17.59(19) & .. \\
		U & 3650 & 81.39(30) & 16.00(02) & 0.89(05) \\
		B & 4450 & 84.12(30) & 16.39(01) & 0.78(03) \\
		V & 5500 & 85.45(30) & 16.27(03) & 0.53(05) \\
		R & 6450 & 85.53(30) & 16.23(02) & 0.55(03) \\
		I &	7870 & 83.56(50) & 16.61(05) & 0.46(05) \\
		\enddata
		\tablenotetext{a}{Uncertainties of peak-light dates, in units of 0.01 day, are 1 $\sigma$.}
		\tablenotetext{b}{Uncertainties of magnitudes, in units of 0.01 mag, are 1 $\sigma$.}
	\end{deluxetable}
	
	\begin{table}	
		\caption{Parameters of SN 2015bq given by SALT2 Model \label{tab:S2_param}}
		\begin{tabular}{ccc}
			\hline 
			\hline 
			Parameter & Value \\
			\hline
			DayMax & 57085.32 $\pm$ 0.04 \\
			Color & 0.05 $\pm$ 0.02 \\
			X1 & 1.61 $\pm$ 0.06 \\
			RestFrameMag\_0\_U & 16.00 $\pm$ 0.05 mag \\
			RestFrameMag\_0\_B & 16.31 $\pm$ 0.02 mag \\
			RestFrameMag\_0\_V & 16.23 $\pm$ 0.01 mag \\
			RestFrameMag\_0\_R & 16.25 $\pm$ 0.02 mag \\
			\hline
		\end{tabular}
	\end{table}

\begin{figure}
		\includegraphics[width=\columnwidth]{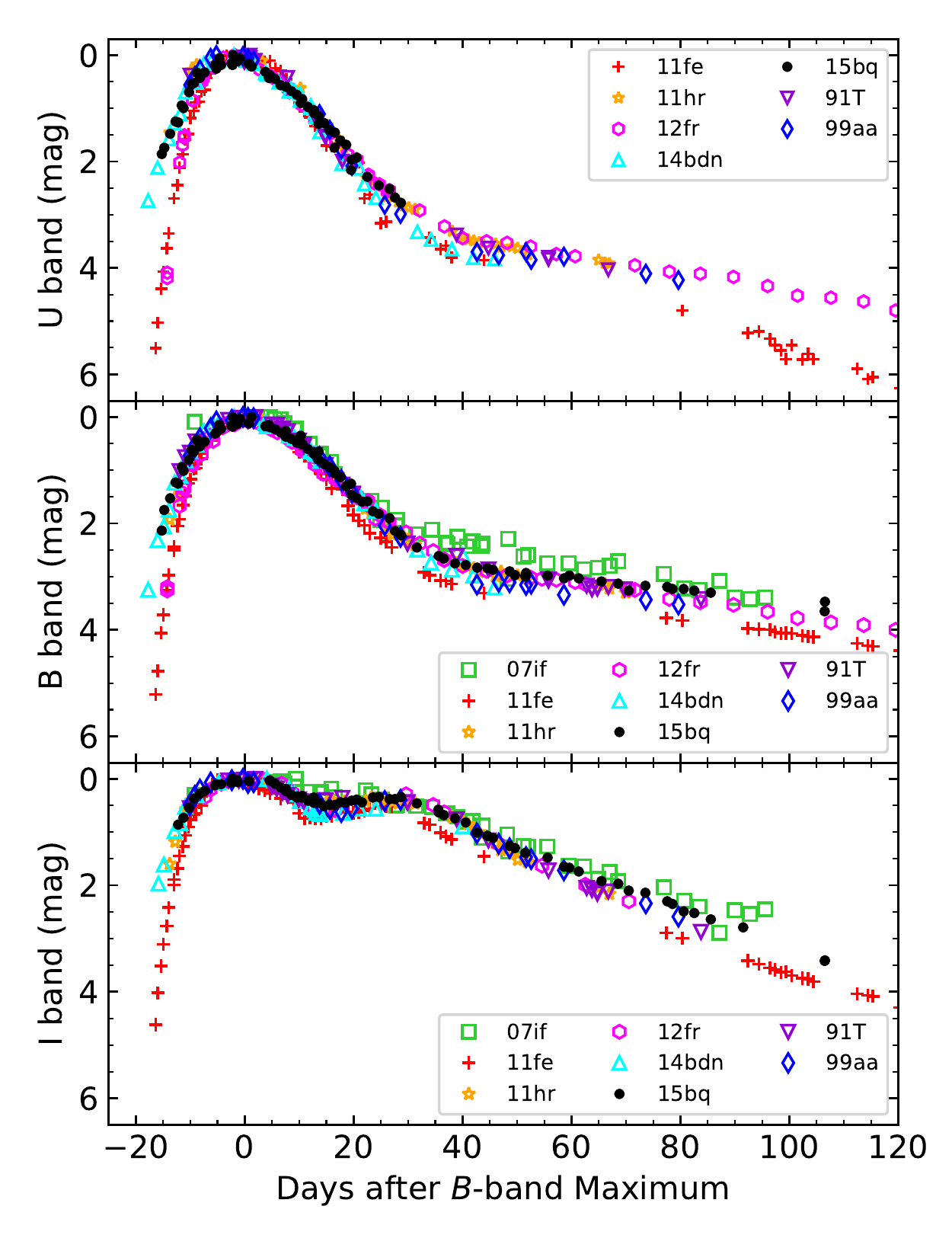}
		\caption{Comparison of the optical light curves of SN 2015bq to other well-observed SNe Ia, including SN 2007if, SN 2011hr, SN 1991T, iPTF 14bdn, SN 1999aa, SN 2012fr, and SN 2011fe; see text for details. \label{fig:LCc}}
	\end{figure}

\begin{figure}
		\includegraphics[width=\columnwidth]{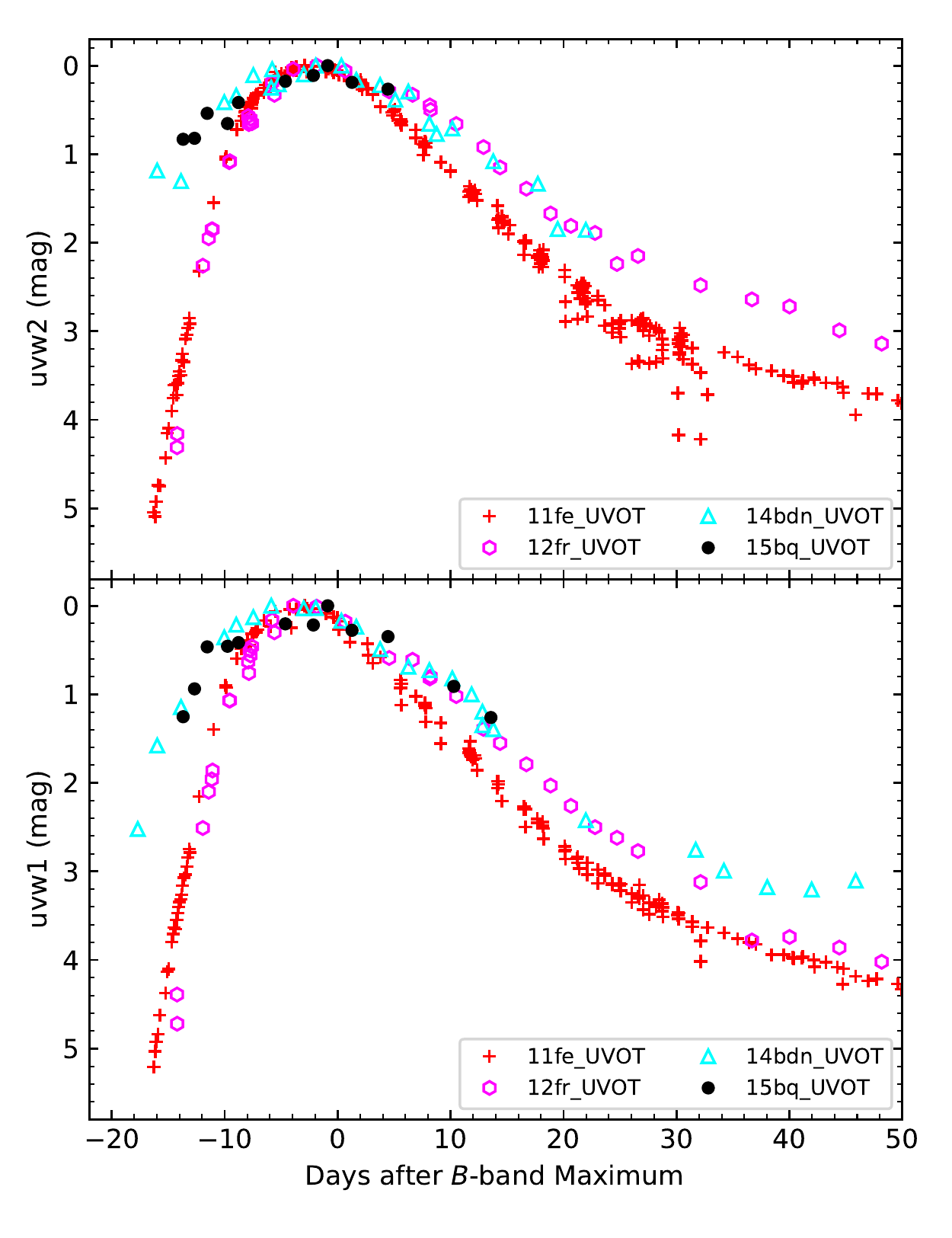}
		\caption{Comparison of the UV light curves of SN 2015bq to SNe Ia with similar decline rates, including  iPTF 14bdn, SN 2012fr, and SN 2011fe; see text for details.\label{fig:UVc}}
	\end{figure}

	Figure \ref{fig:LCc} displays the comparison of SN 2015bq with some well-sampled luminous SS SNe Ia, including the superluminous SN Ia SN 2007if \citep[$\Delta$$m_{15}$ = 0.71 mag;][]{2010ApJ...713.1073S}, luminous SN Ia SN 1991T \citep[$\Delta$$m_{15}$ = 0.95 mag;][]{1998AJ....115..234L}, a transitional object linking 91T-like event to some superluminous SNe Ia SN 2011hr \citep[$\Delta$$m_{15}$ = 0.92 mag;][]{2016ApJ...817..114Z}, the 99aa-like events SN 1999aa \citep[$\Delta$$m_{15}$ = 0.85 mag;][]{2002PhDT........10J} and iPTF 14bdn \citep[$\Delta$$m_{15}$ = 0.84 mag;][]{2015ApJ...813...30S}, and the narrow-lined (NL) SNe Ia 2012fr \citep[$\Delta$$m_{15}$ = 0.85mag;][]{2014AJ....148....1Z}. The normal SN Ia SN 2011fe \citep[$\Delta$$m_{15}$ = 1.08 mag;][]{2012ApJ...753...22B} is also plotted here because of its extremely early and dense observations in multibands.
	The UV-band comparisons are displayed in Figure \ref{fig:UVc}. 
	
	The comparison of light curves shows that SN 2015bq and iPTF 14bdn have higher luminosity than SN 2011fe in the early phase. They are about two magnitudes brighter than the latter one in the UV bands. 
	After the $B$-band maximum, there are no obvious distinctions in \textit{UBI} bands. However, a possible ``shoulder'' is seen in the UV light curves of SN 2015bq, iPTF14bdn, and SN 2012fr. 

	In general, the photometry of SN 2015bq is very similar to iPTF 14bdn in all bands, especially both of them show the excess in the early phase. SN 2015bq and iPTF 14bdn were classified into early-excess SNe Ia (EExSNe) in \citet{2018ApJ...865..149J}. 
	The luminosity enhancement characterizes this kind of SNe Ia in the first few days after the explosions. The origin of the early UV flux will be discussed in Section \ref{sec:origin}.
	
	\subsection{Color Curves and Interstellar Extinction}
\begin{figure*}
		\includegraphics[width=\textwidth]{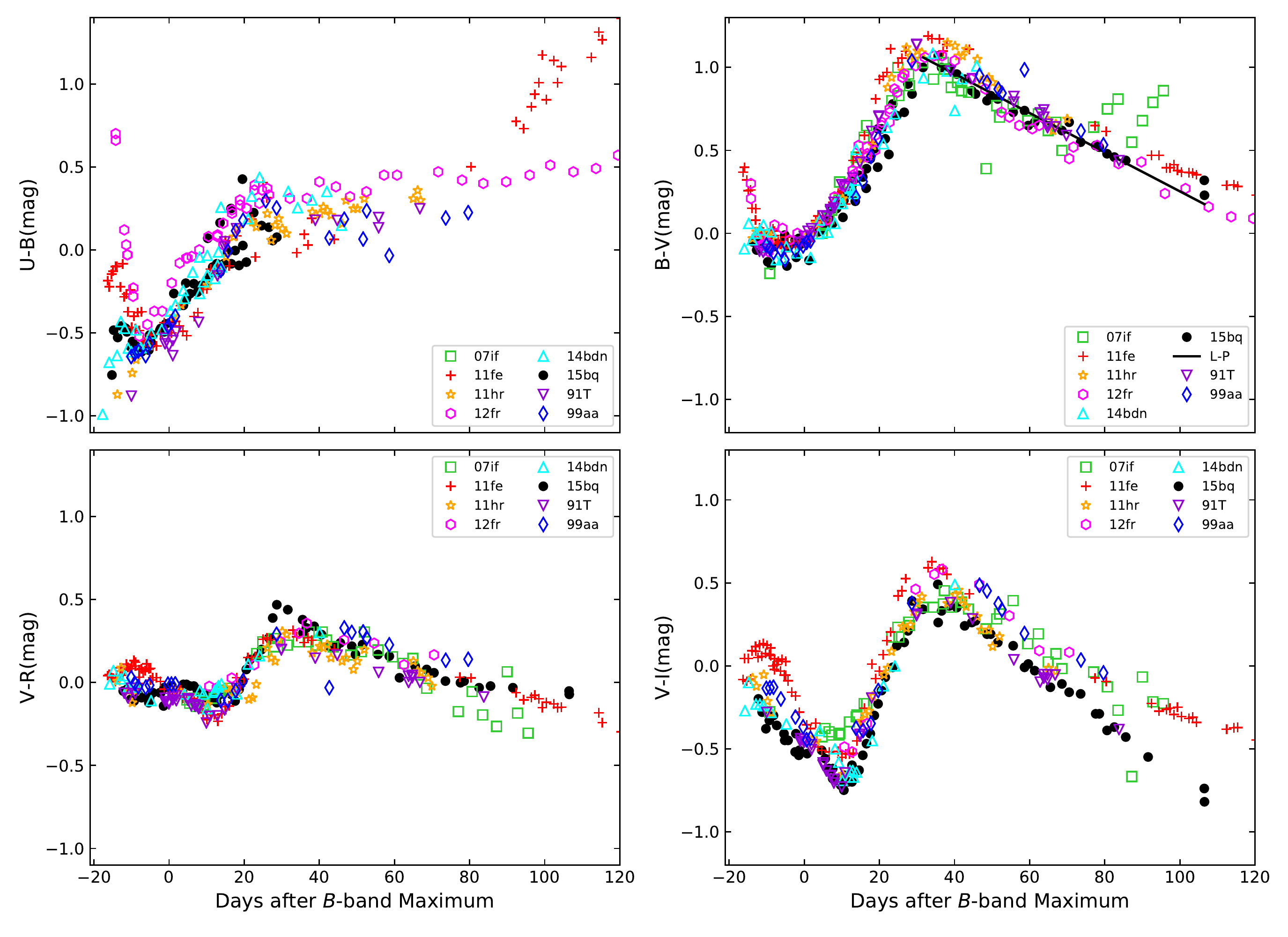}
		\caption{Optical color curves of SN 2015bq compared with SN 2007if, SN 2011hr, SN 1991T, iPTF 14bdn, SN 1999aa, SN 2012fr, and SN 2011fe; see the text for details. \label{fig:CC}}
	\end{figure*}

	Figure \ref{fig:CC} shows the optical color curves of the same sources as Figure \ref{fig:LCc}. All of these color curves are corrected for the reddening of the Milky Way and the host galaxy.  
	The reddening of SN 2015bq, $E(B - V)$ = 0.13 $\pm$ 0.04 mag, is estimated by the intrinsic $B - V$ color assumption of SNe Ia at +30 d $ < $ t $ < $ +90 d \citep[named as Lira-Phillips relation;][]{1999AJ....118.1766P,{2009ApJ...697..380W}}. Considering that the Milky Way reddening presented by the Galaxy dust map is $E(B-V)_{\rm Gal}$ = 0.013 $ \pm $ 0.001 mag \citep{1998ApJ...500..525S}, the reddening due to the host galaxy is $E(B-V)_{\rm host}$ = 0.12 $ \pm $ 0.04 mag. 
	Additionally, the result from the SALT2 fitting gives an extinction coefficient of $E(B - V)$ = 0.15 $\pm$ 0.07 mag, and we adopt the average of the two results, i.e., $E(B - V)$ = 0.14 $\pm$ 0.08 mag.
	
	SN 2015bq is similar to iPTF 14bdn in $B - V$ color, and it seems these luminous SNe Ia are showing bluer color at the early phase.

	At the early phase, 91T-like events show the bluest $U-B$ color among these samples, followed by 99aa-like SNe Ia and normal SNe Ia. Then we noticed that 91T/99aa-like SNe Ia are almost monotonically reddened, while normal SNe Ia shows a distinct ``red-blue-red'' evolution.

	In $V - R$ color, SN 2015bq shows a redder color at about t $ \approx $ +30 d. It might be related to the more significant $R$ band shoulder at this phase. 
	There is a scatter in $V - I$ color, and SN 2015bq shows a blue color in the early phase of $V - I$ color, where the strength of the \ion{Ca}{2} IR triplet could have a dominant effect.
	
	\section{Spectroscopy} \label{sec:4} 
	As shown in Figure \ref{fig:Sp}, the early spectra of SN 2015bq are composed by superior absorption of iron-group elements (IGE) and \ion{Ca}{2} H{$\&$}K, together with the weak absorption of the other IMEs, e.g., \ion{Si}{2} lines, ``W''-shaped \ion{S}{2} lines and \ion{Ca}{2} IR triplet lines. These features class SN 2015bq as a member of the 99aa-like event. More details of spectra evolution are in the following.

\begin{figure*}
	\includegraphics[width=\textwidth]{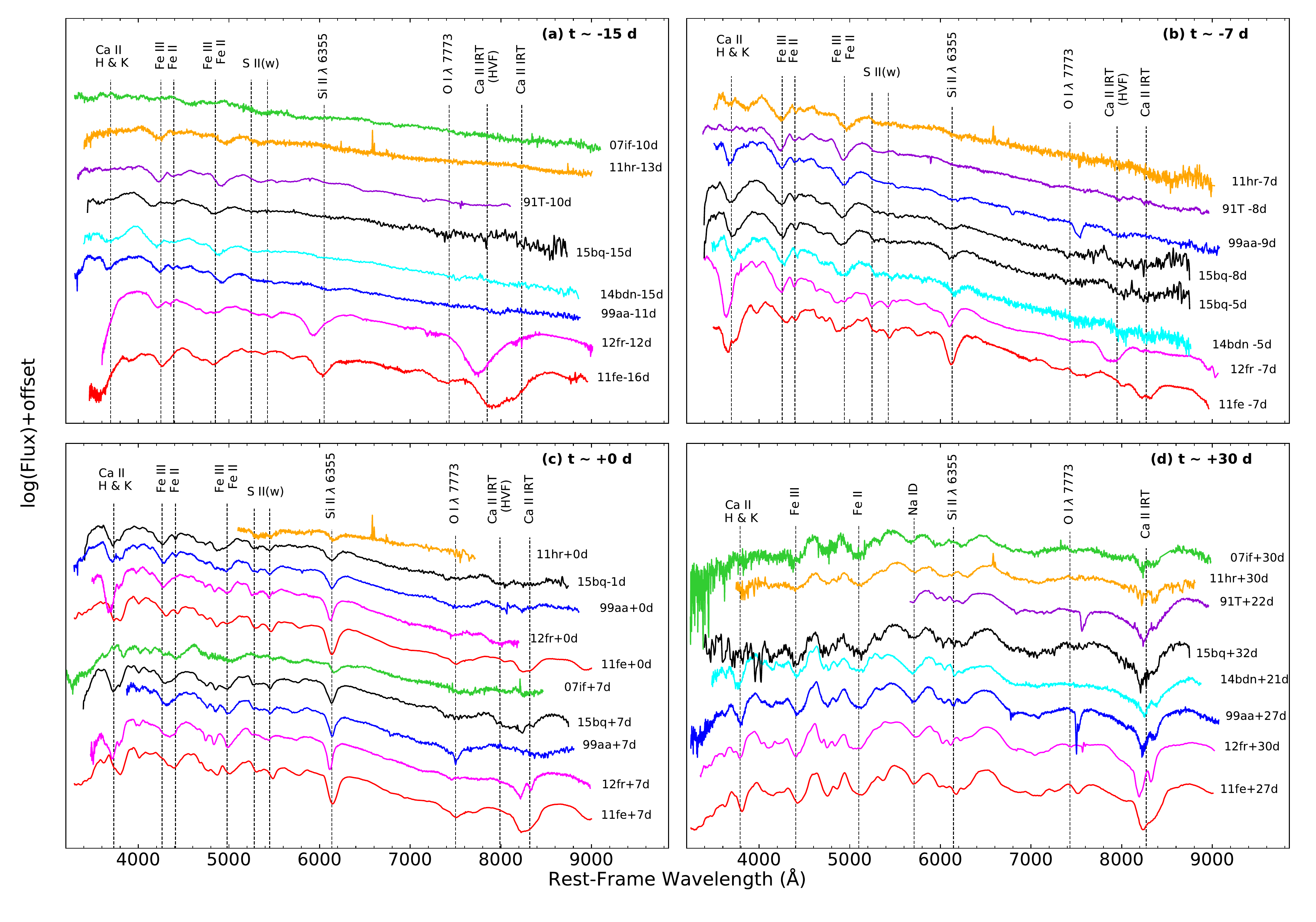}
	\caption{Spectra of SN 2015bq at t $\sim$ $-$15, $-$7, 0, and +30 d after the B-band maximum. Overplotted are the comparable-phase spectra of SN 2007if, SN 2011hr, SN 1991T, iPTF14bdn, SN 1999aa, SN 2012fr, and SN 2011fe.  \label{fig:spec_comp}}
\end{figure*}	
	
	\subsection{Spectra Temporal Evolution} \label{sec:spec}
	We analyze the temporal evolution and comparison of spectra at some selected phases of our sample in this section, including SN 2015bq, SN 2007if, SN 2011hr, SN 1991T, iPTF 14bdn, SN 1999aa, SN 2012fr and SN 2011fe, which could give more details of the explosion.
	
	The early spectra obtained at t $\leq$ $ - $10 d are plotted in Figure \ref{fig:spec_comp}(a). The dominant features at this phase are the high-velocity features \citep[HVFs;][]{2005MNRAS.357..200M,2005ApJ...623L..37M,2015ApJS..220...20Z} of IMEs and high-ionized Fe group elements. At this phase, the spectra of normal SNe Ia show mainly lines of IMEs, including \ion{Si}{2} $\lambda$6355, \ion{Ca}{2} H{$\&$}K, and \ion{Ca}{2} IRT. 
	At around t $\sim$ $-$14 d, the spectrum of SN 2015bq is dominated by the blended absorption lines of \ion{Fe}{3} and \ion{Fe}{2} lines, and the absorption lines of IMEs are almost absent.These are the standard features of 91T/99aa-like events. 
		
	However, the typical 99aa-like events, e.g., SN 1999aa and iPTF 14bdn, show evident absorption of \ion{Ca}{2} H{$\&$}K at this phase, while the 91T-like events do not have such a feature. 
	The spectrum of SN 2015bq has the similar   \ion{Ca}{2} H{$\&$}K  feature, but it is weaker than that of SN 1999aa.
	At t $\sim$ $-$10 d, the spectrum of SN 2007if has no significant feature and is dominated by doubly ionized Fe.  
	The above phenomenon illustrates that the order of ionization degree from high to low is superluminous SNe Ia, 91T-like events, 99aa-like events, and regular events. In addition, this sequence may help probe the temperature, as a higher temperature may lead to fewer IMEs and result in weak IME lines in the spectra \citep{1995A&A...297..509M}.
	
	The large dispersion of the $U - B$ colors in the early phase is closely related to the absorption strength of \ion{Ca}{2} H{$\&$}K and IGEs. Meanwhile, the monotonically reddening $U - B$ color is associated with the strong IGEs lines and the weak IMEs lines. These may be due to the high temperature in the early phase. 
	
	Figure \ref{fig:spec_comp}(b) displays the spectra at t $ \leq $ $-$7 d. At around t $\sim$ $-$5 d, the spectra of SN 2015bq, iPTF 14bdn, and SN 1999aa show a weak absorption of \ion{Si}{2} $\lambda$6355, and the absorption of \ion{Ca}{2} H{$\&$}K becomes stronger. These two  lines are deeper than those of 91T-like events and weaker than those of normal events. 
	The absorption of \ion{S}{2} of SN 2015bq at t $ \sim $ $-$5 d is not as strong as SN 2012fr and SN 2011fe at t $ \sim $ $-$7 d, even weaker than that of  iPTF14bdn at t $ \sim $ $-$5 d.

	The spectral evolution at around the $B$-band maximum is displayed in Figure \ref{fig:spec_comp}(c). At around t $\sim$ 0 d of 99aa-like events, the absorption features of \ion{Si}{2} $\lambda$6355 are becoming more dominant, and \ion{S}{2} lines are becoming noticeable. However, they are still weaker than that in the normal event. 
	At around t $\sim +$7 d, the spectra of SN 2007if begin to show the absorption of IMEs, e.g., \ion{Si}{2} $\lambda$6355, \ion{Ca}{2} H{$\&$}K, and the strength is still weaker than that of 99aa-like events.
	The absorption of \ion{S}{2} of SN 2007if is still not significant.

	The spectral evolution at one month after the $B$-band maximum is displayed in Figure \ref{fig:spec_comp}(d). The spectra are dominated by iron-group elements, \ion{Na}{1}D, \ion{Ca}{2} H{$\&$}K, and a strong \ion{Ca}{2} IR triplet absorption. 
	The double absorption features of \ion{Ca}{2} IR triplet are more prominent in iPTF 14bdn, SN 1999aa, SN 2015bq, and SN 2012fr. At this phase, all the comparison SNe Ia shows a similar spectral evolution.

	The diversity of the spectra of SNe Ia is mainly demonstrated in the early phase.

	\subsection{The Ejecta Velocity}
\begin{figure}
		\includegraphics[width=\columnwidth]{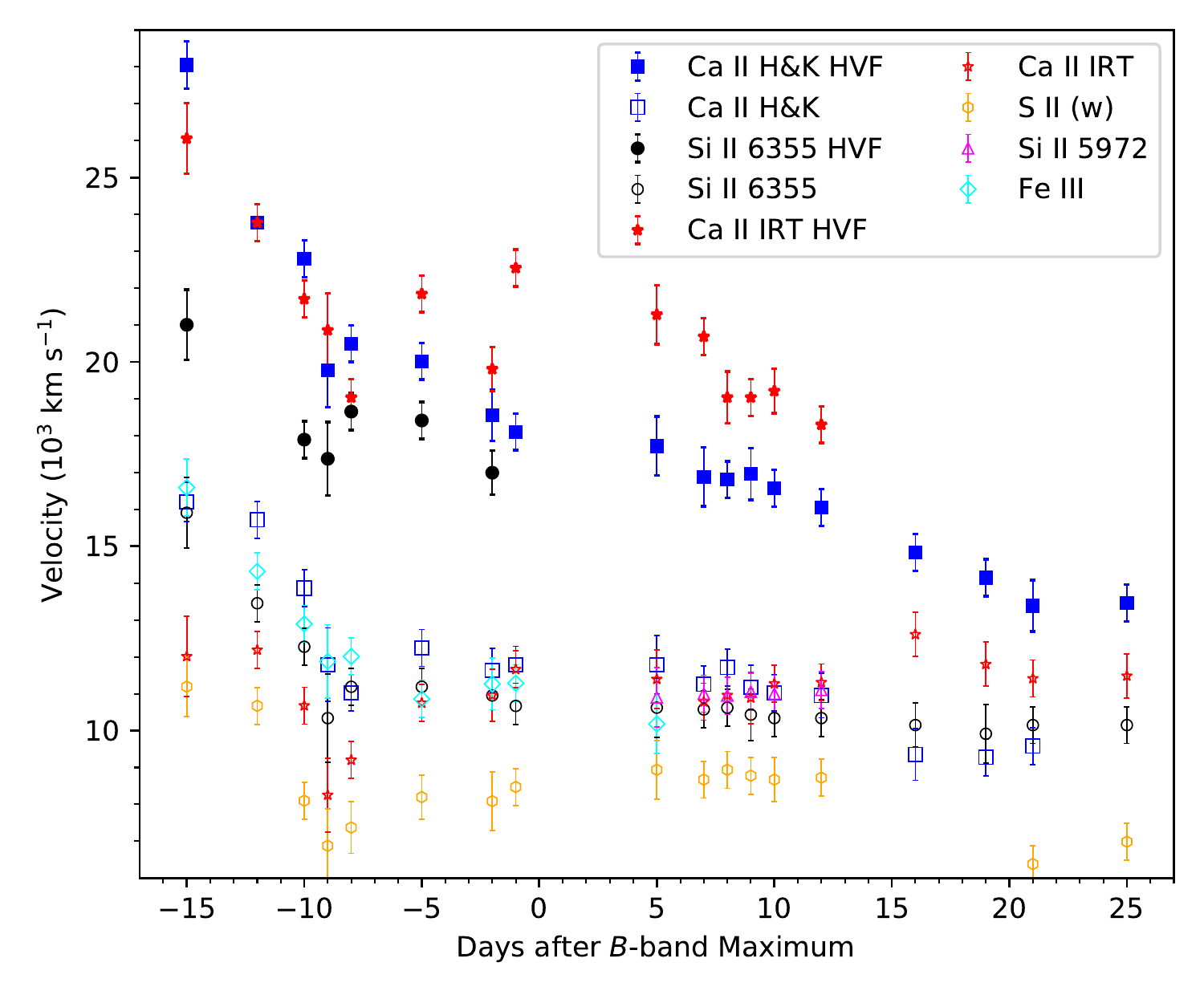}
		\caption{Ejecta velocity evolution of different elements of SN 2015bq.  \label{fig:Vel}}
	\end{figure}

	The ejecta velocities of SN 2015bq derived from the absorption minimum of some spectral lines are displayed in Figure \ref{fig:Vel}, including \ion{Si}{2} $\lambda$6355, \ion{Si}{2} $\lambda$5972, \ion{Ca}{2} H{$\&$}K, \ion{Fe}{3}, \ion{S}{2}(w), \ion{C}{2} $\lambda$6580, \ion{Ca}{2} IR triplet lines, and \ion{Si}{2} $\lambda$6355 HVF, \ion{Ca}{2} H{$\&$}K HVF, \ion{Ca}{2} IRT HVF. 
	The locations of the absorption minimum are measured using both the Gaussian fit routine and direct measurement of the center of the absorption, and the results are averaged.
	
	The velocity of \ion{Fe}{3} at t $\approx$ $-$15 d is $\sim$ 16,000 km $\rm s^{-1}$ and declines quickly, disappeared at t $\approx$ $+$5 d with $\sim$ 11,000 km $\rm s^{-1}$. 
	\ion{Fe}{3} is usually used to estimate the burning efficiency of SNe Ia. The dominated IGEs in the inner extent indicate that SN 2015bq produces more $^{56}{\rm Ni}$ than the normal SNe Ia \citep{2007Sci...315..825M}.

	After the maximum light, the photospheric velocity of \ion{Si}{2} $\lambda$6355, \ion{Ca}{2} H{$\&$}K, and \ion{Ca}{2} IR triplet remains at $ \sim $ 11,000 km $\rm s^{-1}$ for over a month. In contrast, \ion{Si}{2} $\lambda$5972 remains at a similar velocity just for ten days. 
	Thus, the velocity of IMEs seems to have a plateau at t $\approx$ +5 d to t $\approx$ +12 d with a very low velocity gradient, which puts SN 2015bq into the low-velocity gradient (LVG) category of SNe Ia  in the classification scheme of \cite{2005ApJ...623.1011B}. 
	On the other hand, the IMEs of SN 2015bq generally have similar expansion velocities (e.g., $\sim$ 11,000 km $\rm s^{-1}$), indicating that the burning products in the ejecta may have a relatively uniform distribution.

	\section{Discussion}\label{sec:5}
	
	\subsection{The Peak Luminosity and The Nickel Mass} \label{sec:Ni}

	The distance modulus of SN 2015bq derived from the redshift of the host galaxy is $\mu_{0}$ = 35.45 $\pm$ 0.15 mag \citep{2000ApJ...529..786M}, which is close to the result from SALT2 fitting ($\mu_{0}$ = 35.55 $\pm$ 0.14 mag). We adopt the average value of these two estimations and get $\mu_0 = 35.50 \pm 0.15$ mag, D = 126 $\pm$ 9 Mpc.
	
\begin{figure}
		\includegraphics[width=\columnwidth]{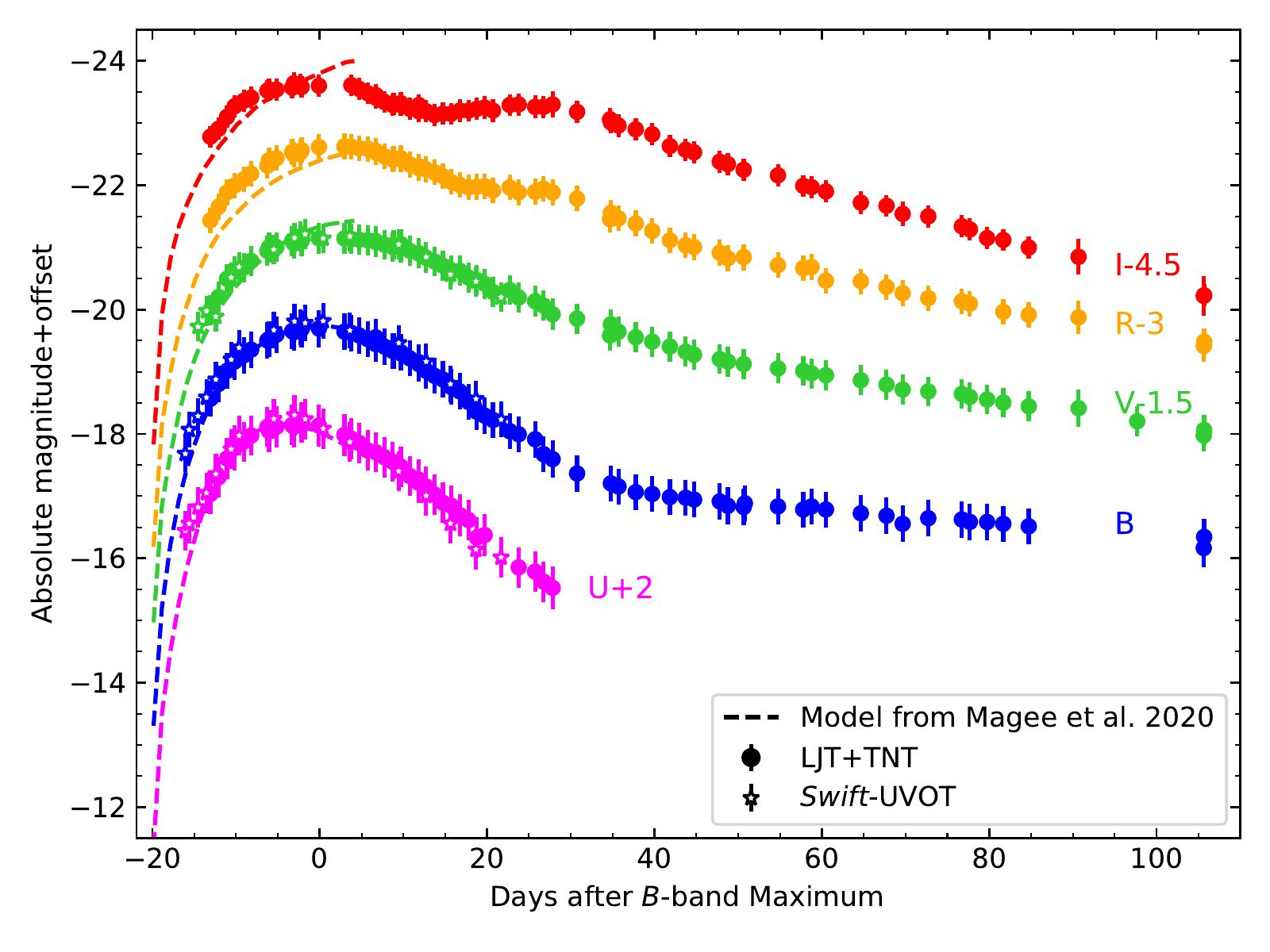}
		\caption{Absolute magnitude light curves of SN 2015bq. The light curves are shifted vertically for better display. The dashed-lines are the best fitting model in \citet{2020A&A...634A..37M}. \label{fig:Magee}}
	\end{figure}

	Adopting this distance and correcting for galactic extinction, we derive that the peak absolute magnitudes of SN 2015bq are $ M_{B} = -19.68 \pm 0.41 $ mag and $ M_{V} = -19.66 \pm 0.37 $ mag. Both are higher than the typical values of SNe Ia \citep[e.g., the typical $ M_{B} = -19.33 \pm 0.06 $ mag and $ M_{V} = -19.27 \pm 0.05 $ mag given by][]{2006ApJ...645..488W}.
	The absolute magnitude light curves are displayed in Figure \ref{fig:Magee}. 
	
	Furthermore, we plot a model, EXP\_Ni0.8\_KE0.50\_P4.4, from TURTLS \citep[a one-dimensional Monte-Carlo radiative transfer code designed for modeling the early time evolution of SNe Ia;][]{2018A&A...614A.115M, 2020A&A...634A..37M} which provides a better fit to the light curves of SN 2015bq than the rest of \cite{2020A&A...634A..37M}.
	
\citet{2020A&A...634A..37M} model produces three parameters of SN 2015bq, where the total $^{56}{\rm Ni}$ mass is 0.8$M_{\sun}$. However, this model has only three options of $^{56}{\rm Ni}$ mass (i.e., 0.4, 0.6, and 0.8$M_{\sun}$), which may not be  accurate enough to draw an actual mass of each particular SN Ia. Therefore, we did not use the result provided by \citet{2020A&A...634A..37M} as the mass of SN 2015bq. 
EXP\_KE0.50 means the exponential density profiles with velocity scale $ v_{e} \sim 1735.91 $ km $\rm s^{-1}$ and kinetic energy $\sim 5.04 \times 10^{50}$ erg. P4.4 is the scaling parameter with a smaller value of 4.4, which gives a more shallow $^{56}{\rm Ni}$ distribution. 
	We also get the rising time from the fitting as $t_{r} = 20.77 \pm 0.31$ d.
	
	However, this model does not fit the peak of light curves at V and I bands. 
	It might be that there are more data points in \textit{UB-}  bands before the \textit{B}-band maximum.
	Because Magee's models simultaneously fit the data points of different bands in the early phase, the wights of \textit{U} and \textit{B} bands with more data are higher.
	
	\begin{figure}
		\includegraphics[width=\columnwidth]{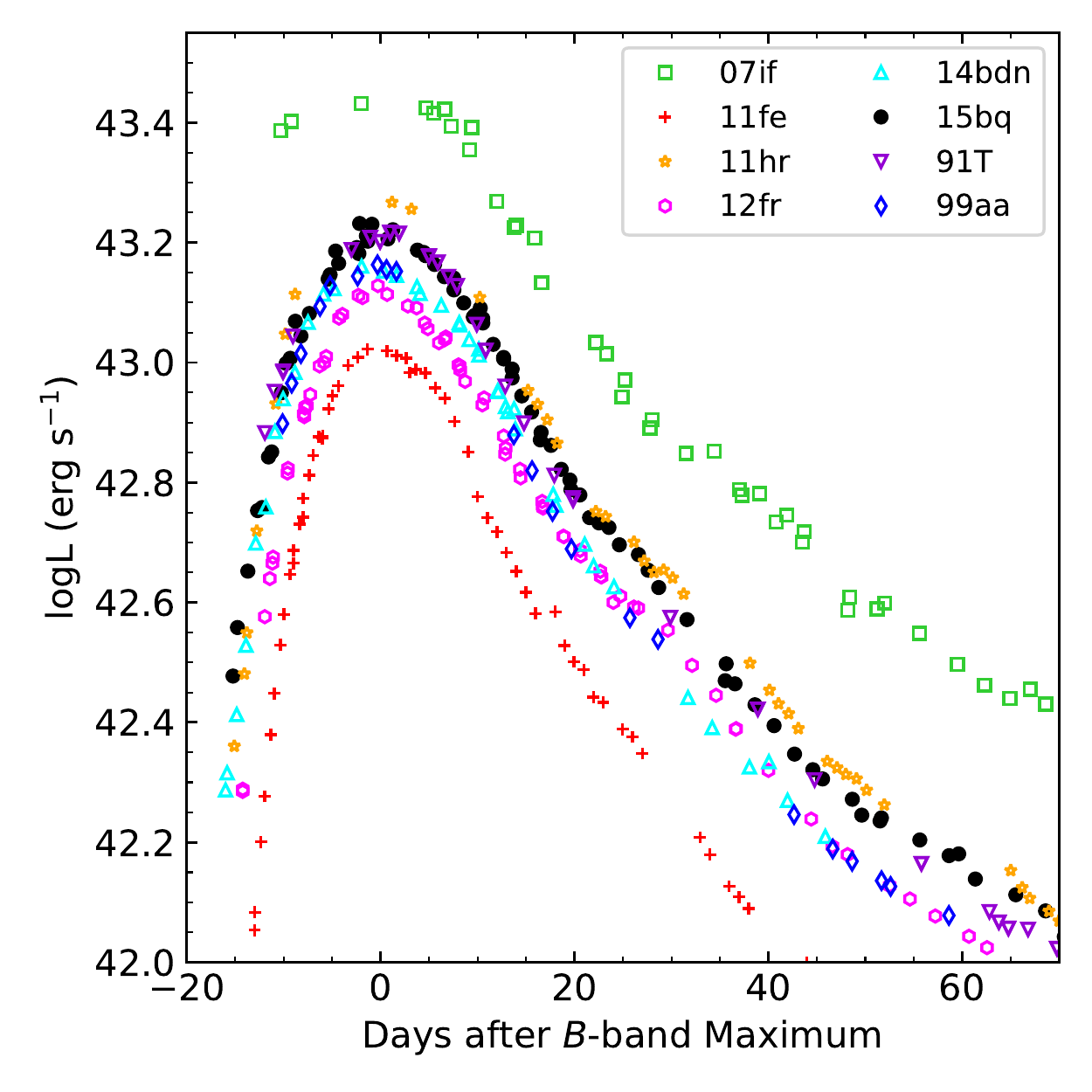}
		\caption{The quasi-bolometric light curves of SN 2015bq derived from \textit{UBVRI} photometry compared with SN 2007if, SN 2011hr, SN 1991T, iPTF 14bdn, SN 1999aa, SN 2012fr, and SN 2011fe. \label{fig:bolo}}
	\end{figure}
	
	\begin{deluxetable*}{ccccccccc}
		\tablecaption{Photometric Parameters \label{tab:pho_param}}
		\tablehead{
			\colhead{name} & \colhead{$t_{r}$} & \colhead{$\Delta m_{15}(B)$} & \colhead{$M_{max}(B)$} & \colhead{$L_{max}$} & \colhead{$M_{ \rm ^{56}{\rm Ni}}$} & \multicolumn{2}{c}{EW(\ion{Si}{2} $\lambda 6355$)} & \colhead {ref}\\
			\cline{7-8}
			\colhead{ } & \colhead{ } & \colhead{ } & \colhead{ } & \colhead{ } & \colhead{ } & \colhead{$\sim 0$ d } & \colhead{ $\sim +7$ d}  &\colhead{ }\\
			\colhead{ } & \colhead{(days)} & \colhead{(mag)} & \colhead{(mag)} & \colhead{($10^{43} \rm erg~s^{-1}$)} & \colhead{($M_{\sun}$)}&\colhead{ (\AA)} &\colhead{(\AA)} & \colhead{}
		}
		\startdata
		SN 2007if & 24.0 & 0.71 & -20.23 & 3.22 & 1.60 & $\cdots$ & 29.95 & (1)\\
		SN 2011hr & 18.0 & 0.92 & -19.84 & 2.30 & 1.11 & 27.74 & $\cdots$ & (2)\\
		SN 1991T & 20.2 & 0.94 & -19.67 & 1.93 & 1.03 & $\cdots$ & 46.28 & (3)\\
		SN 2015bq & 20.77 & 0.82 & -19.68 & 1.75 & 0.97 &43.93& 56.89 & (4)\\
		iPTF 14bdn &  19.9 & 0.84& -19.57 & 1.76 & 0.93	& $\cdots$&$\cdots$ & (5) \\
		SN 1999aa & 19.0 & 0.85 & -19.60 & 1.73 & 0.72 & 55.33& 58.78 & (6)\\
		SN 2012fr & 17.9 & 0.85 & -19.49 & 1.82 & 0.88 & 61.57&61.58 & (7)\\
		SN 2011fe & 17.5 & 1.11 & -19.24 & 1.31 & 0.62 & 79.71 & 88.79 & (8)\\	
		\enddata

		\tablerefs{(1) \citet{2010ApJ...713.1073S}. (2) \citet{2016ApJ...817..114Z}. (3) \citet{1998AJ....115..234L}; \citet{2000AA...359..876C}; \citet{2014MNRAS.445..711S}. (4) This paper.  (5) \citet{2015ApJ...813...30S}; (6) \citet{2004AJ....128..387G,2006AJ....131..527J}. (7) \citet{2013ApJ...770...29C}; \citet{2014AJ....148....1Z}. (8) \citet{2013AA...554A..27P}.}
	\end{deluxetable*}

	The bolometric luminosity of SN 2015bq is the sum of the flux of the four wavelength components: optical band, UV band, near-infrared (NIR) band, and the wavelength shorter than 1600 \AA. 
	The optical flux is calculated by the \textit{UBVRI} photometry.
	Because the optical/UV-band light curves and color curves (see Figure \ref{fig:LCc} - \ref{fig:CC}) of iPTF14bdn are similar to SN 2015bq, we assume these two SNe have the  same uvm2$-$uvw1 color.
	Thus, we derive the missed uvm2-band magnitude of SN 2015bq by its uvw1-band light curve combined with the interpolated uvm2- uvw1 color of iPTF 14bdn.
	The flux of NIR and the wavelength shorter than 1600 \AA\, could be estimated via the measurement of a large sample of SNe Ia.
	In this work, we evaluate that the NIR flux is 5\% of the optical flux around the maximum according to \citet{2009ApJ...697..380W}, and the flux that $<1600$ \AA~ is 3\% of the optical flux according to \cite{2007ApJ...663.1187H}.

	Following the above procedure, the peak bolometric  luminosity of SN 2015bq is calculated to be $ L_{ \rm peak} = (1.75 \pm 0.37) \times 10^{43}~ \rm erg~ s^{-1}$. 

	According to Arnett Law \citep{1982ApJ...253..785A,2005A&A...431..423S} and the derived bolometric luminosity, the synthesized $^{56}{\rm Ni}$ mass in the explosion of SN 2015bq is  $M_{ \rm ^{56}{\rm Ni}} = 0.97 \pm 0.20 M_{\sun}$.

	The major parameters of these SNe Ia derived from the bolometric and $ B $-band light curves are listed in Table \ref{tab:pho_param}. The quasi-bolometric light curves synthesized by \textit{UBVRI} photometry are plotted in Figure \ref{fig:bolo}.

	\subsection{The origin of early-excess flux} \label{sec:origin}
	The early excess seen in the 99aa-like events, such as SN 2015bq (this paper), iPTF 14bdn \citep{2015ApJ...813...30S}, and SN 1999aa \citep{2018ApJ...865..149J}, relate to their dominated \ion{Fe}{3}/\ion{Fe}{2} lines and bluer $U-B$ color at the similar phase.	
	
		There are two physical mechanisms to explain the origin of early-excess SNe Ia: interaction or surface radioactivity. The interaction would happen between ejecta and a nondegenerate companion star \citep{2010ApJ...708.1025K} or dense circumstellar matter \citep[CSM;][]{2016ApJ...826...96P}. Surface radioactivity could be helium detonation \citep[He-det;][]{2018ApJ...861...78M} or surface-$^{56}{\rm Ni}$-decay \citep{2018ApJ...865..149J}. 
	
	SN 2015bq has a small decline rate ($\Delta$$m_{15}$(B) $ \sim 0.82$ mag) and higher peak luminosity ($ L_{ \rm peak} \sim 1.75 \times 10^{43} \rm erg s^{-1}$) than normal SNe Ia, which produces  a more significant amount of Ni ($ M_{ \rm ^{56}{\rm Ni}} \sim 1 M_{\sun}$). 
	It is likely SN 2015bq could have more mass of $^{56}$Ni at the surface than usual, although the fraction of this nickel  may be small.
	Therefore, the surface-$^{56}{\rm Ni}$-decay is preferred to interpret the early excess of SN 2015bq.It is consistent with the surface-$^{56}{\rm Ni}$-decay scenario  \cite{2018ApJ...865..149J} proposed to explains the early excess in 91T/99aa-like events \citep[][]{2019ApJ...872...14Z}. 
	
	The discussion about the $^{56}{\rm Ni}$ mass above is primarily based on a model of ejecta shells outside a radiating photosphere. This model assumes a continuous mass distribution of $^{56}{\rm Ni}$, peaking toward the center of the ejecta \citep[e.g.,][]{1982ApJ...253..785A}. 
	However, the differences between 99aa-like and normal SNe Ia might be due to the off-center distance of the initial ignition point and the view angle effect \citep{2018ApJ...865..149J}. 
	The gravitationally confined detonation \citep[GCD;][]{2004ApJ...612L..37P,2007ApJ...662..459K,2008ApJ...681.1448J} model may describe the explosion of 99aa-like events by adjusting the off-center ignition point. In the GCD scenario, the transition of a deflagration bubble ignited near the stellar core will trigger a detonation at the opposite side, resulting in a more significant amount of $^{56}{\rm Ni}$ at the outer layer.

	Furthermore, \cite{2010ApJ...708.1025K} proposed that the interaction between the expanding ejecta material and the nondegenerate companion star will cause a luminosity enhancement at the early phase and cause a ``bump" feature in the light curve. Nevertheless, it needs a specific view angle, and only $\sim$10$\%$ SNe Ia may show the early excess in the observation. 
	It seems to see a part of ``bump" at early UV light curves of SN 2015bq. 
	However, our observations started not early enough to have the complete early-time light curves.
	Thus we can not make sure there is a ``bump".
	
	In addition, \citet{2018ApJ...865..149J} found that a significant fraction of 91T/99aa-like objects shows early excess. They argued it could not be due to the viewing angle as predicted by the interaction scenario. Thus, the observation of SN 2015bq might prefer surface-$^{56}{\rm Ni}$-decay scenario to interaction channel in terms of early flux excess.

	\subsection{Diversity of the luminous SS SNe Ia} \label{diversity}

Table \ref{tab:pho_param} lists the EWs of \ion{Si}{2} $\lambda$6355 of the spectra presented in Figure \ref{fig:spec_comp}(b). These SNe Ia, except SN 2011fe, can be classified into SS subtype in the Branch diagram \citep{2006PASP..118..560B,2009PASP..121..238B} due to the small EW ($\lesssim$ 60 \AA) of \ion{Si}{2} $\lambda$6355 and the nearly absent \ion{Si}{2} $\lambda$5972 as seen in Figure \ref{fig:spec_comp}(b). We further classify those SNe Ia as luminous SS SNe Ia due to the high peak luminosity listed in this table. 

One can see a sequence of 12fr - 99aa - 15bq - 91T - 11hr - 07if in the profile of \ion{Si}{2} $\lambda$6355 that follows a change from deep and narrow to shallow and wide. Such a tendency corresponds to the luminosity increasing presented in Table \ref{tab:pho_param}. Thus, it seems that the peak luminosity of these SNe Ia decreases with the increasing EW of \ion{Si}{2} $\lambda$6355.  Furthermore, the nickel masses of the explosions are causally varying with the luminosities in different sources, as the decay of nickel is thought to be responsible for the luminosity.

Indeed, SN 2012fr is likely to be a link between normal and 99aa-like \citep{2014AJ....148....1Z}, SN 1999aa as a possible link between 91T-like and normal SNe Ia \citep{2004AJ....128..387G},  SN 2011hr represents a transitional object linking 91T-like SNe Ia to some superluminous SNe Ia \citep{2016ApJ...817..114Z}.
Similarly, SN 2015bq in this work is possibly the connection between the 99aa-like and 91T-like SNe Ia, as the properties of its spectra seem to lie between the two types of SNe Ia (see Figure \ref{fig:spec_comp}). 
In the early phase, a weak absorption of \ion{Ca}{2} H{$\&$}K occurs in the spectrum of SN 2015bq, while there is no corresponding absorption for SN 1991T, and there is a stronger absorption for SN 1999aa. 
Meanwhile, the absorption strength of \ion{Si}{2} $\lambda$6355 of SN 2015bq is likely between 99aa-like and 91T-like SNe Ia (Figure \ref{fig:spec_comp} and Table \ref{tab:pho_param}).
Those comparisons make SN 2015bq a possible intermediate in the ``quasi-evolution'' sequence of the luminous SS SNe Ia.

	\cite{2016AJ....151..125Z} suggested that the luminous SS SNe Ia had small EWs of \ion{Si}{2} $\lambda$6355 could be derived from different profiles of this line with smaller depth and larger width or with larger depth and small width. 
	In the Branch and Wang diagram \citep{2009PASP..121..238B,2009ApJ...699L.139W}, as shown in the left panel of Figure \ref{fig:SS}, SN 2015bq is located at the left bottom corner, which is considered the luminous SS SNe Ia region. 
	SNe Ia in this region shows different properties that can be classified into another subclass, such as SC, 91T-like, 99aa-like, and NL SNe Ia.
    In the right panel, SN 2015bq is between the 91T-like SNe Ia and NL SNe Ia. It is consistent with the sequence mentioned above.

    Although the spectra of those SNe Ia are not identical in the same phase (see Figure \ref{fig:spec_comp}), the distinct structures in the spectra may be attributed to the variation of temperatures, as discussed in Section \ref{sec:spec}.
    Thus, the temperature might be a dominant factor in the luminous SS SNe Ia sequence from high to normal luminosity. 
    It might suggest that those SNe originate from a similar mechanism. 
    Such a scenario can naturally explain the possible tendency of the parameters in Table \ref{tab:pho_param}, e.g., the smaller EW of the absorption line of \ion{Si}{2} $\lambda$6355, the more thoroughly the silicon burns and the greater the luminosity.

\begin{figure}
		\includegraphics[width=\columnwidth]{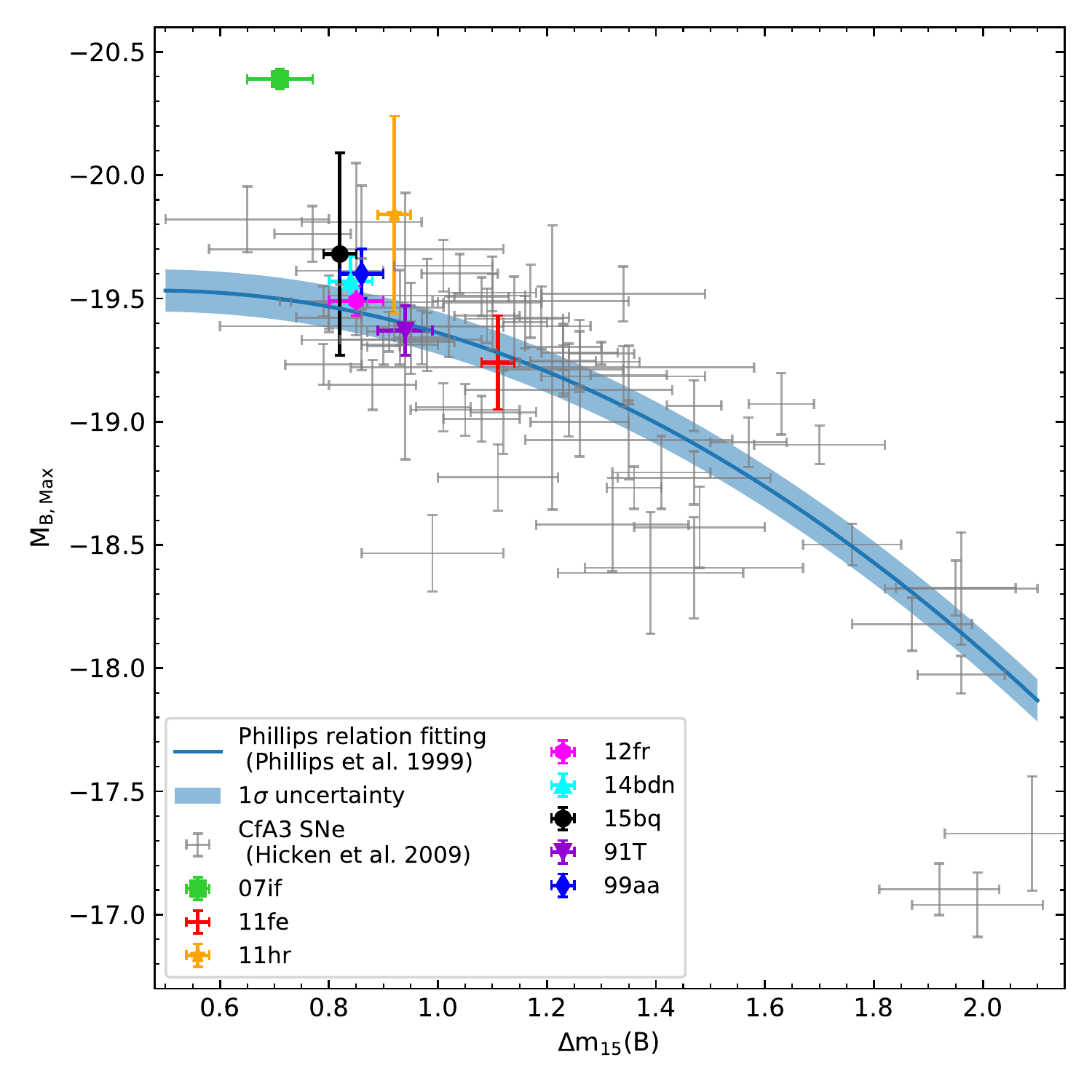}
		\caption{Absolute-magnitude [$ M_{max}(B) $] vs. decline-rate [$\Delta m_{15}(B)$] diagram for SN 2007if, SN 2011hr, SN 1991T, SN 2015bq, iPTF 14bdn, SN 1999aa, SN 2012fr and SN 2011fe, and a subsample of the CfA3 SNe Ia \citep{2009ApJ...700..331H}. The solid curve is the best-fit relation proposed by  \citet{1999AJ....118.1766P} for the CfA3 data. The blue band represents the 1$\sigma$ uncertainty of the Phillips relation fitting. \label{fig:cfa}}
	\end{figure}

\begin{figure*}
		\includegraphics[width=\textwidth]{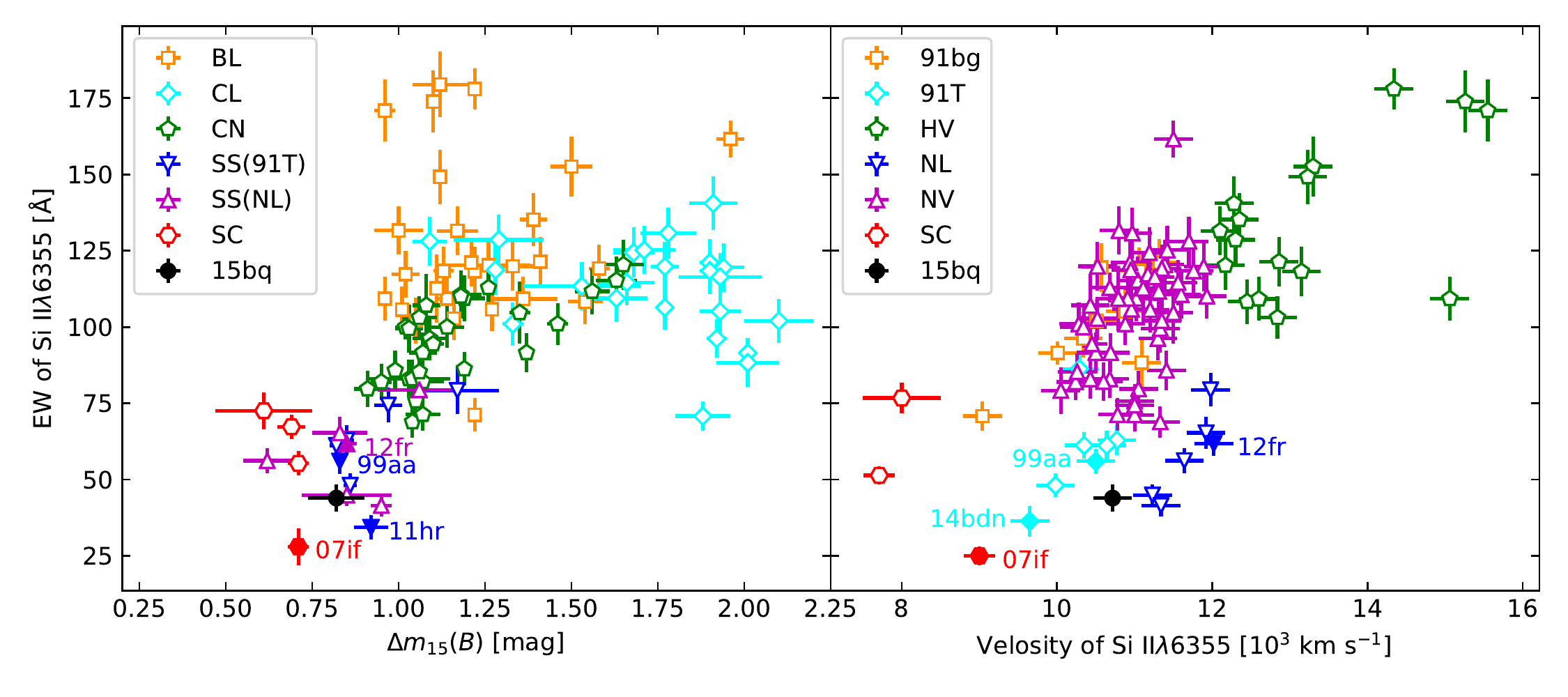}
		\caption{Comparison of various spectroscopic and photometric indicators from SN 2015bq with those from other SNe Ia as measured by \citet{2012AJ....143..126B}, \citet{2012MNRAS.425.1917S}, \citet{2009ApJ...699L.139W}, \citet{2014AJ....148....1Z}, \citet{2016AJ....151..125Z} and this paper. The selected sample have spectra within t$\pm $3 d. The left panel is $\Delta m_{15}(B)$ vs. the EW of \ion{Si}{2} $\lambda$6355 at maximum light with subclasses defined by \citet{2009PASP..121..238B} and \citet{2016AJ....151..125Z}; The right panel is the velocity of \ion{Si}{2} $\lambda$6355 vs. the EW of \ion{Si}{2} $\lambda$6355 at maximum light with subclasses defined by \citet{2009ApJ...699L.139W} and \citet{2014AJ....148....1Z}. \label{fig:SS}}
	\end{figure*}

	To check whether those luminous SS SNe Ia are fitted in the prediction of WLR or not, we make a comparison with the CfA3 sample \citep{2009ApJ...700..331H}.  As shown in Figure \ref{fig:cfa}, most of them are still located at the WLR prediction with the  CfA3 sample considering the measurement error, except SN 2007if.  
	It indicates that the luminosity differences of those luminous SS SNe Ia are concentrated in the early phase, and their properties tend to be the same near the maximum.
	Based on the discussion in Section \ref{sec:origin}, the $^{56}{\rm Ni}$-abundance outer layer in the early phase might cause a variety of luminous SS SNe Ia.

	\section{Conclusion}\label{sec:6}
	In general, SN 2015bq is a luminous SN Ia with a slow rise and decline rate in the light curve. It reached the apparent peak magnitude of $m_{\rm max}(B) = 16.39\pm0.01$ mag and an absolute peak magnitude of $M_{\rm max}(B) = -19.68\pm0.41$ mag. Using UVBRI-band light curves, we derive a peak luminosity of $ L_{ \rm peak} = (1.75 \pm 0.37) \times 10^{43}~ \rm erg~ s^{-1}$ for SN 2015bq, corresponding to a synthesized nickel mass of $ M_{ \rm ^{56}{\rm Ni}} = 0.97 \pm 0.20 M_{\sun}$. 
	The light-curve evolution is similar to that of another 1999aa-like event like iPTF 14bdn.
	The $U-B$ color is bluer than the normal SNe Ia and became monotonically reddened early. 
	In particular, prominent excess emission can be seen in the UV bands. 
	Meanwhile, its early spectra are characterized by prominent features of IGEs and \ion{Ca}{2} H{$\&$}K absorptions, along with relatively weak absorption of IMEs, e.g., the shallow absorption features of \ion{Si}{2} $\lambda$6355 and \ion{Ca}{2} IRT, and there is no obvious difference compared to normal SNe Ia after maximum light.

	The surface-$^{56}{\rm Ni}$-decay might be responsible for the early excess found in the UV-band light curves of SN 2015bq since the derived mass of synthesized $^{56}{\rm Ni}$ in SN 2015bq is more significant than that in normal SNe Ia.

	It seems that there is an evolutionary sequence in the parameters (e.g., $\Delta m_{15}(\rm{B})$, $M_{max}(B)$, $U - B$ color, $ M_{ \rm ^{56}{\rm Ni}}$, the EW of \ion{Si}{2} $\lambda$6355) of the luminous SS SNe Ia.
	With the spectral comparisons, SN 2015bq seems to be an inbetween object linking 91T-like and 99aa-like subclasses.
	The variations intrinsic to the sequence may be related to the temperature, which indicates that the origin of those luminous SS SNe Ia may be similar.
    The difference between them only exists in the early phase, which may be caused by different amounts of nickel mixed into the outer layers.

\section*{Acknowledgements}
We thank  the anonymous referee for the constructive comments and suggestions.
	We acknowledge the support of the staff of the LJT, TNT and XLT. Funding for the LJT has been provided by the CAS and the People's Government of Yunnan Province. The LJT is jointly operated and administrated by YNAO and Center for Astronomical Mega-Science, CAS. JZ is supported by the National Natural Science Foundation of China (NSFC, grants 11773067, 12173082, 11403096), by the Youth Innovation Promotion Association of the CAS (grant 2018081), and by the Ten Thousand Talents Program of Yunnan for Top-notch Young Talents.  XW is supported by NSFC (grants 12033003, and 11761141001), the Major State Basic Research Development Program (grant 2016YFA0400803), and the Scholar Program of Beijing Academy of Science and Technology (DZ:BS202002).  This work is partially supported by National key research and development program 2018YFA0404204, the Science Foundation of Yunnan Province (No. 2018FA004).  We acknowledge the science research grants from the China Manned Space Project with NO. CMS-CSST-2021-A13.

	\bibliography{15bq}{}
	\bibliographystyle{aasjournal}
	
	
	
	\appendix
	\section{Photometry and spectroscopy data of SN 2015bq}
	\setcounter{table}{0} 
\renewcommand{\thetable}{\Alph{section}\arabic{table}}

	\begin{deluxetable}{cccccc}[h]
		\tablecaption{Local Photometric Standards in the \textit{UBVRI} bands \label{tab:stand}}
		\tablehead{
			\colhead{Star} & \colhead{$U$ (mag)} & \colhead{$B$ (mag)} & \colhead{$V$ (mag)} & \colhead{$R$ (mag)} &\colhead{$I$ (mag)}
		}
		\startdata
		1 & 18.10(04) & 18.16(03) & 17.49(02) & 17.10(02) & 16.74(04) \\
		2 & 14.46(02) & 14.52(01) & 14.23(02) & 14.18(01) & 14.08(03) \\
		3 & 17.39(04) & 17.70(02) & 17.22(03) & 16.99(02) & 16.71(04) \\
		4 & 16.27(02) & 15.26(04) & 14.05(02) & 13.37(02) & 12.71(02) \\
		5 & 17.90(04) & 18.22(02) & 17.68(02) & 17.34(04) & 16.99(02) \\
		6 & 15.39(02) & 15.14(02) & 14.38(03) & 13.97(01) & 13.57(02) \\
		7 & 18.41(02) & 18.14(04) & 17.31(02) & 16.80(01) & 16.33(02) \\
		8 & 16.57(02) & 15.53(02) & 14.15(02) & 13.13(01) & 12.12(03) \\
		9 & 17.99(05) & 17.17(03) & 15.80(04) & 14.77(02) & 13.72(04) \\
		10 & 18.46(05) & 17.70(03) & 16.65(03) & 15.97(01) & 15.51(04) \\
		\enddata	
		\tablecomments{See Figure \ref{fig:img} for the finder chart of SN 2015bq and the comparison stars. Uncertainties, in units of 0.01 mag, are 1$\sigma$.}
	\end{deluxetable}
	
	\startlongtable
	\begin{deluxetable*}{cccccccc}
		\tablecaption{Optical Photometry of SN 2015bq at LJT \label{tab:Opti_Pho}}
		\tablehead{
			\colhead{MJD} & \colhead{Epoch\tablenotemark{a}} & \colhead{U (mag)} & \colhead{B (mag)} & \colhead{V (mag)} & \colhead{R (mag)} & \colhead{I (mag)} & \colhead{Telescope}
		}
		\startdata
		57071.91 & -12.21 & 17.13(03) &	17.51(03)	&	17.44(03)	&	17.39(02)	&	17.43(02)	&	LJT	\\
		57072.9	& -11.22	&	16.86(08)	&	17.27(03)	&	17.23(03)	&	17.17(03)	&	17.30(03)	&	LJT	\\
		57073.9	& -10.22	&	16.56(07)	&	17.06(03)	&	16.94(03)	&	16.94(02)	&	17.11(03)	&	LJT	\\
		57074.81 & -9.31	&	16.39(16)	&	16.91(04)	&	16.82(04)	&	16.83(03)	&	16.94(03) &	LJT	\\
		57075.91 & -8.21	&	16.29(01)	&	16.81(03)	&	16.76(03)	&	16.73(01)	&	16.85(02)	& LJT	\\
		57076.78 & -7.34 &	16.19(02)	&	16.72(03)	&	16.65(03)	&	16.64(02)	&	16.80(02)	&	LJT	\\
		57078.72 & -5.4	&	16.05(02)	&	16.57(03)	&	16.49(03) &	16.51(02)	&	16.69(02)	&	LJT	\\
		57078.92 & -5.2	&	16.12(11)	&	16.55(04)	&	16.43(04)	&	16.42(02)	&	16.67(03)	&	LJT	\\
		57079.8	& -4.32	&	16.05(05) &	16.48(03)	&	16.43(03)	&	16.39(02)	&	16.67(02)	&	LJT	\\
		57081.68 & -2.44	&	16.02(03)	&	16.42(03)	&	16.33(03)	&	16.29(03)	&	16.64(03)	&	LJT	\\
		57081.89 &	-2.23	&	16.05(06)	&	16.44(04)	&	16.37(04)	&	16.33(02)	&	16.57(03)	&	LJT	\\
		57082.66 &	-1.46 & 	$\cdots$	 &	16.36(04)	&	16.26(04)	&	16.30(05)	&	16.59(03)	&	TNT \\	
		57082.81 &	-1.31	&	15.98(04)	&	16.39(05)	&	16.33(05)	&	16.26(03)	&	16.63(02)	&	LJT \\	
		57084.88 &	0.76	&	16.03(10)	&	16.38(08)	&	16.29(08)	&	16.21(04)	&	16.61(03)	&	LJT \\	
		57087.93 &	3.81 &	16.18(11)	&	16.43(07)	&	16.29(07)	&	16.20(04)	&	$\cdots$	&	LJT \\	
		57088.75 &	4.63	&	16.24(07)	&	16.45(05)	&	16.30(05)	&	16.21(02)	&	16.60(02)	&	LJT \\	
		57089.69 &	5.57	&	16.31(03)	&	16.49(04)	&	16.31(04)	&	16.23(03)	&	16.66(03)	&	LJT \\	
		57090.71 &	6.59	&	16.43(03)	&	16.55(03)	&	16.32(03)	&	16.24(02)	&	16.73(02)	&	LJT \\	
		57091.66 &	7.54	&	$\cdots$	&	16.52(04)	&	16.33(03)	&	16.27(03)	&	16.80(04) &	TNT \\	
		57091.69 &	7.57	&	16.46(03)	&	16.63(03)	&	16.35(03)	&	16.30(02)	&	16.76(02)	&	LJT \\	
		57092.7	&	8.58	&	16.54(03)	&	16.67(02)	&	16.39(02)	&	16.36(03)	&	16.87(02)	&	LJT \\	
		57093.69 &	9.57	&	16.61(02)	&	16.74(02)	&	16.43(02)	&	16.41(01)	&	16.92(02)	&	LJT	\\
		57093.82	&	9.7 &	$\cdots$ &	16.74(02)	&	16.39(03)	&	16.40(03)	&	16.90(04)	&	TNT \\	
		57094.66	&	10.54	&	$\cdots$ &	16.76(03)	&	16.38(02)	&	16.39(04)	&	16.92(03)	&	TNT \\	
		57094.67 &	10.55	&	16.69(05)	&	16.77(02)	&	16.44(02)	&	16.41(02)	&	16.89(03)	&	LJT \\	
		57095.74 &	11.62	&	16.84(04)	&	16.86(02)	&	16.49(02)	&	16.49(05)	&	16.98(02)	&	LJT \\	
		57096.81 &	12.69	&	16.90(07)	&	16.92(03)	&	16.53(03)	&	16.55(01)	&	17.02(02)	&	LJT \\	
		57096.82 &	12.7	&	16.95(04)	&	16.95(02)	&	16.53(02)	&	16.52(01)	&	16.92(01)	&	LJT \\	
		57097.69 &	13.57	&	17.01(03)	&	17.06(02)	&	16.60(02)	&	16.57(01)	&	17.02(02)	&	LJT \\	
		57098.69 &	14.57	&	17.14(02)	&	17.14(01)	&	16.66(01)	&	16.62(01)	&	17.08(01)	&	LJT \\	
		57099.7	&	15.58	&	17.27(06)	&	17.20(05)	&	16.73(05)	&	16.68(03)	&	17.06(03)	&	LJT	\\
		57100.68 &	16.56	&	17.32(06)	&	17.32(02)	&	16.80(02)	&	16.77(02)	&	17.06(02)	&	LJT	\\
		57101.7	&	17.58	&	17.47(09)	&	17.39(05)	&	16.81(05)	&	16.82(01)	&	17.01(05)	&	LJT	\\
		57102.76 &	18.64	&	17.55(06)	&	17.56(03)	&	16.92(03)	&	16.86(02)	&	17.01(03)	&	LJT	\\
		57103.76 &	19.64	&	17.83(05)	&	17.72(06)	&	16.96(06)	&	16.85(03)	&	16.98(03)	&	LJT	\\
		57104.68 &	20.56	&	17.79(09)	&	17.78(04)	&	17.02(04)	&	16.84(02)	&	16.96(05)	&	LJT	\\
		57105.68 &	21.56 &	$\cdots$	&	17.85(03)	&	17.15(04)	&	16.91(02)	&	17.01(04)	&	TNT	\\
		57107.68 &	23.56	&	$\cdots$	&	18.03(02)	&	17.12(02)	&	16.86(02)	&	16.92(02)	&	LJT	\\
		57108.75 &	24.63	&	18.31(05)	&	18.08(04)	&	17.24(04)	&	16.94(02)	&	16.91(04) &	LJT	\\
		57110.72 &	26.6	&	18.38(06)	&	18.16(06)	&	17.30(06)	&	16.93(04)	&	16.95(06)	&	LJT	\\
		57111.72 &	27.6	&	18.54(05)	&	18.40(04)	&	17.37(04)	&	16.88(02)	&	16.95(04)	&	LJT	\\
		57112.81 &	28.69	&	18.64(12)	&	18.48(08)	&	17.51(08)	&	16.94(04)	&	16.91(12)	&	LJT	\\
		57115.73 &	31.61	&	$\cdots$	&	18.71(03)	&	17.58(03)	&	17.04(02)	&	17.03(03)	&	LJT	\\
		57119.67 &	35.55	&	$\cdots$	&	$\cdots$	&	17.85(07)	&	17.37(07)	&	17.15(05)	&	TNT	\\
		57119.77 &	35.65	&	$\cdots$	&	18.87(03)	&	17.67(03)	&	17.27(02)	&	17.20(04)	&	LJT	\\
		57120.68 &	36.56	&	$\cdots$	&	18.92(03)	&	17.79(03)	&	17.36(01)	&	17.25(05)	&	LJT	\\
		57122.74 &	38.62	&	$\cdots$	&	19.01(01)	&	17.88(01)	&	17.44(03)	&	17.31(02)	&	LJT	\\
		57124.71 &	40.59	&	$\cdots$	&	19.04(04)	&	17.95(04)	&	17.56(04)	&	17.39(02)	&	LJT	\\
		57126.82 &	42.7	&	$\cdots$	&	19.09(04)	&	18.03(04)	&	17.71(03)	&	17.58(03)	&	LJT	\\
		57128.71 &	44.59	&	$\cdots$	&	19.10(03)	&	18.11(03)	&	17.79(02)	&	17.64(02)	&	LJT	\\
		57129.71 &	45.59	&	$\cdots$	&	19.13(02)	&	18.16(02)	&	17.82(01)	&	17.68(01)	&	LJT	\\
		57132.78 &	48.66	&	$\cdots$	&	19.16(03)	&	18.23(02)	&	17.91(01)	&	17.83(02)	&	LJT	\\
		57133.75 &	49.63	&	$\cdots$	&	19.23(07)	&	18.27(04)	&	18.00(05)	&	17.87(05)	&	TNT	\\
		57135.64 &	51.52	&	$\cdots$	&	19.25(06)	&	18.31(04)	&	17.98(05)	&	17.96(05)	&	TNT	\\
		57135.78 &	51.66	&	$\cdots$	&	19.19(04)	&	$\cdots$	&	$\cdots$	&	$\cdots$	&	LJT	\\
		57139.74 &	55.62	&	$\cdots$	&	19.24(04)	&	18.38(04)	&	18.11(01)	&	18.05(04)	&	LJT	\\
		57142.76 &	58.64	&	$\cdots$	&	19.29(03)	&	18.42(04)	&	18.16(03)	&	18.22(02)	&	LJT	\\
		57143.75 &	59.63	&	$\cdots$	&	19.24(03)	&	18.46(03)	&	18.14(02)	&	18.24(04)	&	LJT	\\
		57145.47 &	61.35	&	$\cdots$	&	19.29(02)	&	18.49(03)	&	18.36(01)	&	18.31(03)	&	LJT	\\
		57149.64 &	65.52	&	$\cdots$	&	19.35(02)	&	18.57(02)	&	18.37(01)	&	18.49(05)	&	LJT	\\
		57152.71 &	68.59	&	$\cdots$	&	19.39(02)	&	18.64(02)	&	18.46(01)	&	18.54(02)	&	LJT	\\
		57154.64 &	70.52	&	$\cdots$	&	19.52(06)	&	18.72(06)	&	18.56(07)	&	18.67(10)	&	TNT	\\
		57157.7	&	73.58	&	$\cdots$	&	19.43(03)	&	18.75(02)	&	18.64(01)	&	18.71(03)	&	LJT	\\
		57161.71 &	77.59	&	$\cdots$	&	19.45(03)	&	18.79(03)	&	18.69(02)	&	18.87(04)	&	LJT	\\
		57162.67 &	78.55	&	$\cdots$	&	19.49(02)	&	18.84(03)	&	18.73(01)	&	18.92(04)	&	LJT	\\
		57164.73 &	80.61	&	$\cdots$	&	19.49(03)	&	18.88(02)	&	$\cdots$	&	19.06(05)	&	LJT	\\
		57166.67 &	82.55	&	$\cdots$	&	19.52(02)	&	18.93(03)	&	18.86(02)	&	19.09(03)	&	LJT	\\
		57169.71 &	85.59	&	$\cdots$	&	19.56(02)	&	18.99(02)	&	18.91(02)	&	19.21(03)	&	LJT	\\
		57175.66 &	91.54	&	$\cdots$	&	$\cdots$	&	19.02(18)	&	18.95(18)	&	19.36(23)	&	TNT	\\
		57182.67 &	98.55	&	$\cdots$	&	$\cdots$	&	19.23(03)	&	$\cdots$	&	$\cdots$	&	LJT	\\
		57190.62 &	106.5	&	$\cdots$	&	19.91(09)	&	19.46(08)	&	19.41(15)	&	19.99(26)	&	TNT	\\
		57190.67 &	106.55	&	$\cdots$	&	19.73(04)	&	19.37(03)	&	19.34(03)	&	19.98(05)	&	LJT	\\
		\enddata
		\tablecomments{Uncertainties, in units of 0.01 mag, are 1$\sigma$; MJD = JD - 2400000.5.}
		\tablenotetext{a}{Referring to the peak of $B$ band on March 03.12 2015, JD. 2457084.62.}
	\end{deluxetable*}
	
	\begin{deluxetable*}{cccccccc}
		\tablecaption{$Swift$ UVOT Photometry of SN 2015bq \label{tab:UVOT}}
		\tablehead{
			\colhead{MJD} & \colhead{Epoch\tablenotemark{a}} & \colhead{$uvw2$ (mag)} &\colhead{$uvm2$ (mag)\tablenotemark{b}} & \colhead{$uvw1$ (mag)} & \colhead{$U$ (mag)} & \colhead{$B$ (mag)} & \colhead{$V$ (mag)} 
		}
		\startdata
		57068.9104 & -15.71 & $\cdots$ & 19.57 & $\cdots$ & 17.72(15) & 18.39(18) & $\cdots$ \\
		57069.3702 & -15.25 & $\cdots$ & 19.60 & $\cdots$ & 17.61(14) & 18.01(14) & $\cdots$ \\
		57070.4379 & -14.18 & 19.62(26) & 19.70 & 18.83(30) & 17.34(11) & 17.79(11) & 17.71(24) \\
		57071.4439 & -13.17 & 19.61(26) & 19.64 & 18.52(22) & 17.11(10) & 17.49(10) & 17.46(19) \\
		57072.5678 & -12.05 & 19.33(30) & 18.97 & 18.05(22) & 16.81(12) & 17.20(11) & 17.55(26) \\
		57074.3692 & -10.25 & 19.45(25) & 19.59 & 18.04(16) & 16.41(07) & 16.88(07) & 16.92(14) \\
		57075.3376 & -9.28 & 19.21(23) & 19.57 & 18.00(17) & 16.19(08) & 16.69(08) & 16.75(13) \\
		57079.4904 & -5.13 & 18.97(18) & 19.64 & 17.79(14) & 15.92(07) & 16.40(07) & 16.47(10) \\
		57081.967 & -2.65 & 18.90(21) & 19.46 & 17.80(17) & 15.86(08) & 16.27(07) & 16.28(11) \\
		57083.2353 & -1.38 & 18.80(26) & 19.06 & 17.59(19) & 15.93(10) & 16.29(09) & 16.22(14) \\
		57085.3893 & 0.77 & 18.98(17) & 19.66 & 17.86(14) & 16.08(07) & 16.26(06) & 16.29(10) \\
		57088.5759 & 3.96 & 19.06(18) & 19.70 & 17.93(14) & 16.29(07) & 16.41(06) & 16.26(09) \\
		57094.4055 & 9.79 & $\cdots$ & $\cdots$ & 18.50(21) & 16.77(08) & 16.61(07) & 16.39(10) \\
		57097.6882 & 13.07 & $\cdots$ & $\cdots$ & 18.85(28) & 17.16(10) & 16.91(07) & 16.59(11) \\
		57100.5955 & 15.98 & $\cdots$ & $\cdots$ & $\cdots$ & 17.60(13) & 17.27(09) & 16.87(13) \\
		57103.6473 & 19.03 & $\cdots$ & $\cdots$ & $\cdots$ & 18.02(18) & 17.51(10) & 16.98(15) \\
		57106.6416 & 22.02 & $\cdots$ & $\cdots$ & $\cdots$ & 18.15(21) & 17.84(13) & 17.24(19)	 \\
		\enddata
		\tablecomments{Uncertainties, in units of 0.01 mag, are 1$\sigma$; MJD = JD - 2400000.5.}
		\tablenotetext{a}{Referring to the peak of $B$ band on March 03.12 2015, JD. 2457084.62.}
		\tablenotetext{b}{The uvm2 magnitudes are 3$\sigma$ limit.}
	\end{deluxetable*}
	
	\begin{deluxetable*}{cccccccc}
		\tablecaption{Spectroscopic observation journal of SN 2015bq \label{tab:spec}}
		\tablehead{
			\colhead{UT Date} & \colhead{MJD} & \colhead{Epopch\tablenotemark{a}} & \colhead{Res} & \colhead{range} & \colhead{Exp. Time} & \colhead{Airmass} & \colhead{Telescope}\\
			\colhead{ } & \colhead{ } & \colhead{days} & \colhead{(\AA~$\rm{pixel}^{-1}$)} & \colhead{(\AA)} & \colhead{(s)} & \colhead{ } & \colhead{(+Instrument)}
		}
		\startdata
		Feb. 18 & 57071.93 & -12.19 & 18 & 3500-9100 & 1800 & 1.16 & LJT YFSOC \\
		Feb. 20 & 57073.90 & -10.22 & 18 & 3500-9100 & 1800 & 1.09 & LJT YFSOC \\
		Feb. 21 & 57074.84 & -9.28 & 18 & 3500-9100 & 1800 & 1.00 & LJT YFSOC \\
		Feb. 22 & 57075.91 & -8.21 & 18 & 3500-9100 & 2700 & 1.14 & LJT YFSOC \\
		Feb. 25 & 57078.89 & -5.23 & 18 & 3500-9100 & 1800 & 1.10 & LJT YFSOC \\
		Feb. 28 & 57081.86 & -2.26 & 18 & 3500-8700 & 1800 & 1.05 & LJT YFSOC \\
		Mar. 1 & 57082.77 & -1.35 & 18 & 3500-9100 & 2400 & 1.02 & LJT YFSOC \\
		Mar. 7 & 57088.72 & 4.60 & 18 & 3500-9100 & 2400 & 1.09 & LJT YFSOC \\
		Mar. 9 & 57090.72 & 6.60 & 18 & 3500-9100 & 2400 & 1.07 & LJT YFSOC \\
		Mar. 10 & 57091.70 & 7.58 & 18 & 3500-9100 & 2400 & 1.10 & LJT YFSOC \\
		Mar. 11 & 57092.72 & 8.60 & 18 & 3500-9100 & 2400 & 1.06 & LJT YFSOC \\
		Mar. 12 & 57093.69 & 9.57 & 18 & 3500-9100 & 2400 & 1.14 & LJT YFSOC \\
		Mar. 14 & 57095.71 & 11.59 & 18 & 3500-9100 & 2400 & 1.06 & LJT YFSOC \\
		Mar. 18 & 57099.71 & 15.59 & 18 & 3500-9100 & 2400 & 1.04 & LJT YFSOC \\
		Mar. 21 & 57102.72 & 18.60 & 18 & 3500-9100 & 2700 & 1.01 & LJT YFSOC \\
		Mar. 24 & 57105.61 & 21.49 & 27 & 3800-8800 & 2700 & 1.07 & XLT BFSOC \\
		Mar. 28 & 57108.94 & 24.82 & 18 & 3500-9100 & 1520 & 1.00 & LJT YFSOC \\
		Apr. 3 & 57115.75 & 31.63 & 18 & 3500-9100 & 2700 & 1.02 & LJT YFSOC \\
		Apr. 8 & 57120.72 & 36.60 & 18 & 3500-9100 & 3000 & 1.01 & LJT YFSOC \\
		Apr. 14 & 57126.83 & 42.71 & 18 & 3600-9100 & 3000 & 1.44 & LJT YFSOC \\
		Apr. 20 & 57132.79 & 48.67 & 18 & 3500-9100 & 3000 & 1.27 & LJT YFSOC \\
		May 10 & 57152.75 & 68.63 & 18 & 3620-9100 & 1500 & 1.39 & LJT YFSOC \\
		May 11 & 57153.75 & 69.63 & 49 & 3770-8700 & 2400 & 1.40 & LJT YFSOC \\
		May 22 & 57164.67 & 80.55 & 49 & 3600-8750 & 1415 & 1.14 & LJT YFSOC \\
		\enddata
		\tablenotetext{a}{Referring to the peak of $B$ band on March 03.12 2015, JD. 2457084.62.}
	\end{deluxetable*}

\end{document}